\documentclass[aps,prd,eqsecnum,nofootinbib,twocolumn,floatfix,superscriptaddress]{revtex4}
\usepackage{graphicx}
\usepackage{times,mathptm}
\newcommand{\beq}{\begin{equation}}
\newcommand{\eeq}{\end{equation}}
\newcommand{\bea}{\begin{eqnarray}}
\newcommand{\eea}{\end{eqnarray}}

\renewcommand{\d}{\delta}
\renewcommand{\l}{\lambda}

\renewcommand{\b}{\beta}

\newcommand{\p}{\phi}

\newcommand{\g}{\gamma}

\newcommand{\m}{\mu}
\renewcommand{\r}{\rho}

\newcommand{\s}{\sigma}

\newcommand{\D}{{\Delta}}

\newcommand{\E}{{\cal E}}

\newcommand{\J}{{\cal J}}

\newcommand{\vph}{\varphi}
\newcommand{\oh}{{\textstyle{\frac{1}{2}}}}

\newcommand{\dg}{\dagger}
\newcommand{\non}{\nonumber}

\newcommand{\rf}[1]{(\ref{#1})}
\newcommand{\ra}{\rightarrow}

\newcommand{\FIGURE}[2][v]{\begin{figure}[#1]#2
              \end{figure}}
\begin{document}
%
%
\title{Localized eigenmodes of covariant Laplacians in the Yang-Mills vacuum}

\author{J. Greensite}
\affiliation{Physics and Astronomy Dept., San Francisco State
University, San Francisco, CA~94117, USA}
\author{{\v S}. Olejn\'{\i}k}
\affiliation{Institute of Physics, Slovak Academy
of Sciences, SK--845 11 Bratislava, Slovakia}
\author{M.I. Polikarpov}
\author{S.N. Syritsyn}
\affiliation{Institute of Theoretical and Experimental Physics,
B. Cheremushkinskaya 25, Moscow 117259, Russia}

\author{V.I. Zakharov}
\affiliation{Max-Planck Instit\"ut f\"ur Physik,
F\"ohringer Ring 6, D-80805 Munich, Germany}
\date{\today}
\begin{abstract}
      As a probe of the Yang-Mills vacuum, we study numerically
the eigenmode spectrum of the covariant lattice Laplacian
operator.  We find that the eigenmodes at the low and high ends of
the spectrum are localized in finite regions whose volume is
insensitive to the lattice volume. We also find that the vacuum is
seen very differently by localized modes of the covariant
Laplacian in different representations of the gauge group.  In the
fundamental representation, the data suggests that the
localization volume is finite in physical units set by the string
tension, and localization disappears when center vortices are
removed.  In the adjoint and $j=3/2$ representations the low and
high-lying modes are far more localized, and the localization
volume appears to scale to zero, in physical units, in the
continuum limit.  The adjoint Laplacian is insensitive to vortex
removal, but we find that exponential localization is absent for
adjoint eigenmodes in the Higgs phase of a gauge-Higgs theory.
Localization is also absent in the spectrum of the Coulomb gauge
Faddeev-Popov operator, as required in Coulomb gauge confinement
scenarios.

\end{abstract}

\pacs{11.15.Ha, 12.38.Aw}
\keywords{Confinement, Lattice Gauge Field Theories, Solitons
Monopoles and Instantons}
\maketitle
%
%
\section{Introduction}\label{Introduction}

    The zero and low-lying eigenmodes of the lattice Dirac operator
are of considerable interest in lattice QCD, because of the
connection of low-lying modes (via the Banks-Casher relation
\cite{Banks}) to the chiral condensate, and the connection of zero
modes  (via the Atiyah-Singer index theorem \cite{Atiyah}) to
topological charge. It has been shown recently by Golterman and
Shamir \cite{Maarten}, for the two-flavor Wilson-Dirac operator
outside the Aoki phase, that the low-lying eigenmodes are
localized in finite regions which do not grow with lattice volume.
The phenomenon is clearly related to Anderson localization
\cite{Anderson} of particle energy eigenstates in a disordered
medium. On the other hand, Aubin et al.\ \cite{MILC} find that the
zero modes of the Asqtad \cite{Asqtad} lattice Dirac operator
extend throughout the lattice, and yet concentrate in regions
which scale with coupling like manifolds of effective dimension
between two and three.\footnote{In this connection, we should also
take note of the rather different results obtained by the Kentucky
group \cite{Ivan}, who compute topological charge density via the
overlap operator.  Their finding is that most of the lattice
volume is taken up by two space-filling folded sheets of opposite
topological charge density, and dimensionality $D=3$.} It is
conceivable that this finding relates to the suggestion
\cite{Valya} that thin brane-like objects must exist in the QCD
vacuum, a suggestion which is supported by studies of the action
density at P-vortex locations \cite{ITEP}.

   It would be interesting to know whether the localization (or
concentration) of low-lying eigenmodes is peculiar to lattice
Dirac operators, or whether the localization feature is generic in
confining lattice gauge theories, and could perhaps serve as
another type of order parameter for confinement.  To address this
question, we study in this article the localization properties of
the spectrum of the covariant lattice Laplacian in the
fundamental, adjoint, and $j=3/2$ representations of the SU(2)
gauge group. Unlike the Dirac operator, the covariant Laplacian is
straightforward to construct on the lattice, having neither chiral
symmetries which ought to be respected, nor doublers which need to
be removed.

   Our main finding will be that there are indeed localized eigenmodes
of the covariant Laplacian at the extreme lower and upper ends of
the spectrum, and we will argue below that localization implies
that neither the physical mass of scalar particles, nor the mass
of scalar particle bound states, can be adjusted to zero via the
bare mass term, at least in the quenched approximation. However, a
very unexpected result of our investigation concerns the
qualitative dependence of localization on the gauge group
representation of the covariant Laplacian.  In the fundamental
$j=\oh$ representation of color SU(2), localized states appear to
concentrate in 4-volumes of finite extension in physical units,
and this localization disappears when center vortices are removed
from the gauge field configurations.  In sharp contrast, in the
adjoint and $j=3/2$ representation, the low-lying eigenmodes are
localized in far smaller regions, whose volume seems to go to zero
in physical units in the continuum limit.

  We also find that localization is absent in the low-lying
eigenmodes of the Coulomb gauge Faddeev-Popov operator, despite
that operator's close resemblance to the covariant Laplacian. This
absence of localization is implied by the existence of a confining
color Coulomb potential, and is therefore required in Coulomb
gauge confinement scenarios.

  In section 2, below, we present our results for the Laplacian
eigenmodes in the fundamental representation.  Section 3 is
concerned with the eigenmodes of the adjoint representation
Laplacian, in both pure Yang-Mills and gauge-Higgs theory. Results
for the $j=3/2$ representation are presented in section 4. Section
5 concerns the relation between localized modes and the absence of
massless particle states; this section also contains our results
for the Coulomb gauge Faddeev-Popov operator.  Our conclusions are
in section 6.

\section{Localization in the Fundamental Representation}

    The covariant lattice Laplacian in the fundamental representation
of the gauge group is given by
\beq
      \triangle_{xy}^{ab} = \sum_{\m} \Bigl[ U^{ab}_\m(x) \d_{y,x+\hat{\m}}
         + U^{\dg ab}_\m(x-\hat{\m}) \d_{y,x-\hat{\m}}  - 2 \d^{ab} \d_{xy} \Bigl]
\label{triangle}
\eeq
(color indices $a,b$ for SU(2) are 1-2)
and we are interested in the low-lying eigenmodes $\p^a_n(x)$ satisfying the
eigenvalue equation
\beq
         -\triangle_{xy}^{ab} \phi_n^b(y) = \l_n \phi_n^a(x)
\eeq
Both the eigenmodes and eigenvalues are, of course, dependent on
the configuration $U_\m(x)$. The observables relevant for
localization are the Inverse Participation Ratios (IPR), defined as
\beq
          IPR_n = V \langle \sum_x \r_n^2(x) \rangle
\eeq
where $\r_n(x)$ is the normalized eigenmode density
\beq
         \r_n(x) = \sum_a \Bigl| \phi^a_n(x) \Bigr|^2
\eeq
with $\sum_x \r_n(x)=1$, and $V=\sum_x 1$ is the lattice volume
in lattice units.
If the n-th eigenmode is localized, with $\rho(x) \sim 1/b$
in some volume $b$ and $\rho(x) \approx 0$ elsewhere, then the IPR
will grow linearly with lattice volume, on a hypercubic lattice
of length $L$, as
\beq
        IPR_n = {L^4 \over b}
\eeq
In contrast, if the eigenmode is extended over the full
lattice with $\rho(x) \sim 1/L^4$, then the IPR will not grow
with volume.

   Of course, intermediate situations in which the eigenmode
is partially localized are also possible. It could be, for
example, that while the eigenmode is largely concentrated in some
region of volume $b$, still there is a substantial fraction of the
norm which is accounted for by the bulk of the lattice volume.
For this reason it is useful to study the falloff in $\rho(x)$
away from its region of maximum concentration.  We introduce the
"Remaining Norm" (RN) observable, defined for a given eigenmode as
follows:  Place the values of $\r_n(x)$ into a 1-dimensional array
$r(i),~  i=1,2,..,V$, sorted so that $r(i) \ge r(i+1)$ .  Then the
remaining norm is given by
\beq
          R(K) = 1 - \sum_{i=1}^K r(i)
\label{RN}
\eeq
The RN is the amount of the total norm ($=1$) remaining after
counting contributions from the subset of  $K < V$ sites having
the largest values of $\r(x)$.   If the curve $R(K)$ becomes more
extended as the lattice volume increases, without approaching
some limiting envelope, then the eigenmode is
not really localized in a fixed volume around a single site.
It should be noted that a closely related quantity,
\beq
       Q(x) = \sum_{i=1}^K r(i)
\eeq
where $x=K/V$ is the volume fraction, was introduced previously
by Horv\'ath in the second article of ref.\ \cite{Ivan}, where it is called
the ``normalized cumulative function of intensity."

\FIGURE[tb]{
\centerline{{\includegraphics[width=8truecm]{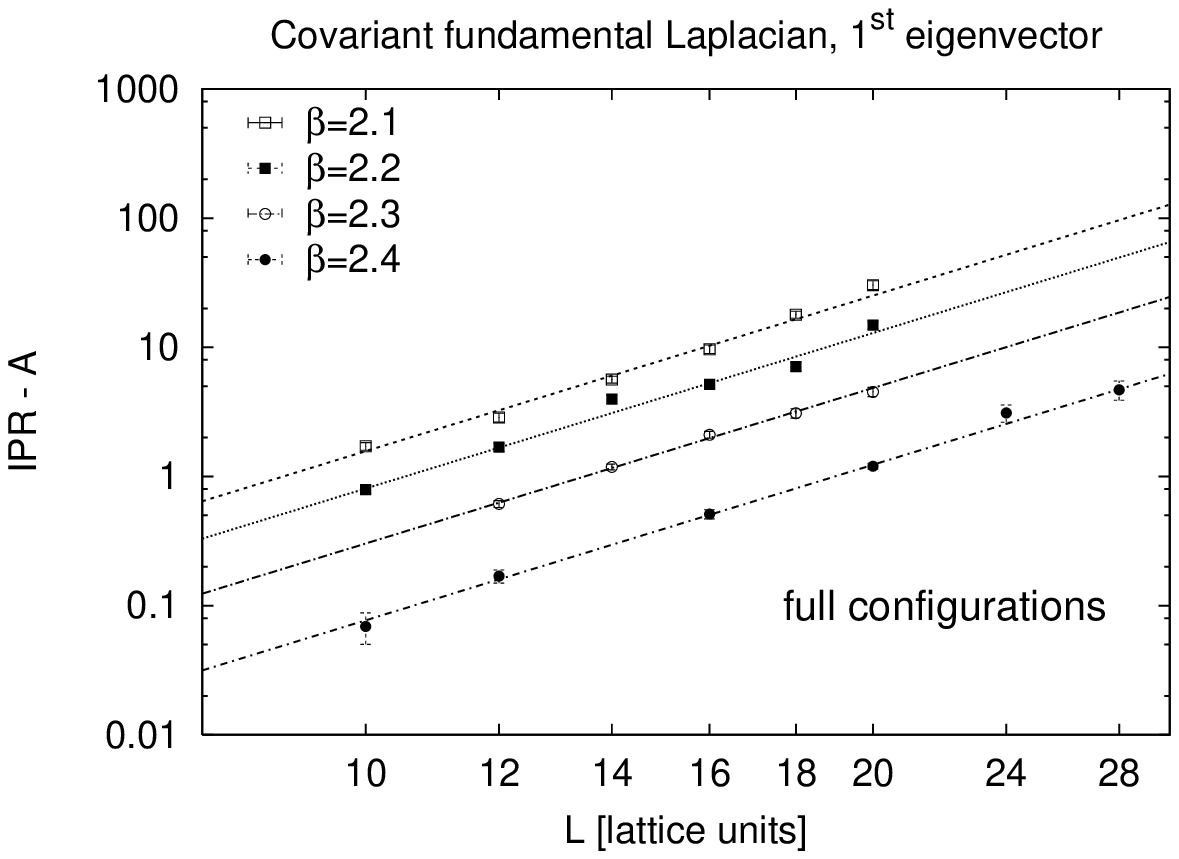}}}
\caption{Log-log plot of $IPR-A$ vs.\ lattice length $L$, for the
lowest eigenmode of the covariant Laplacian in the fundamental
representation, at $\b=2.1-2.4$.} \label{IPR-fund} }

\FIGURE[tb]{
\centerline{{\includegraphics[width=8truecm]{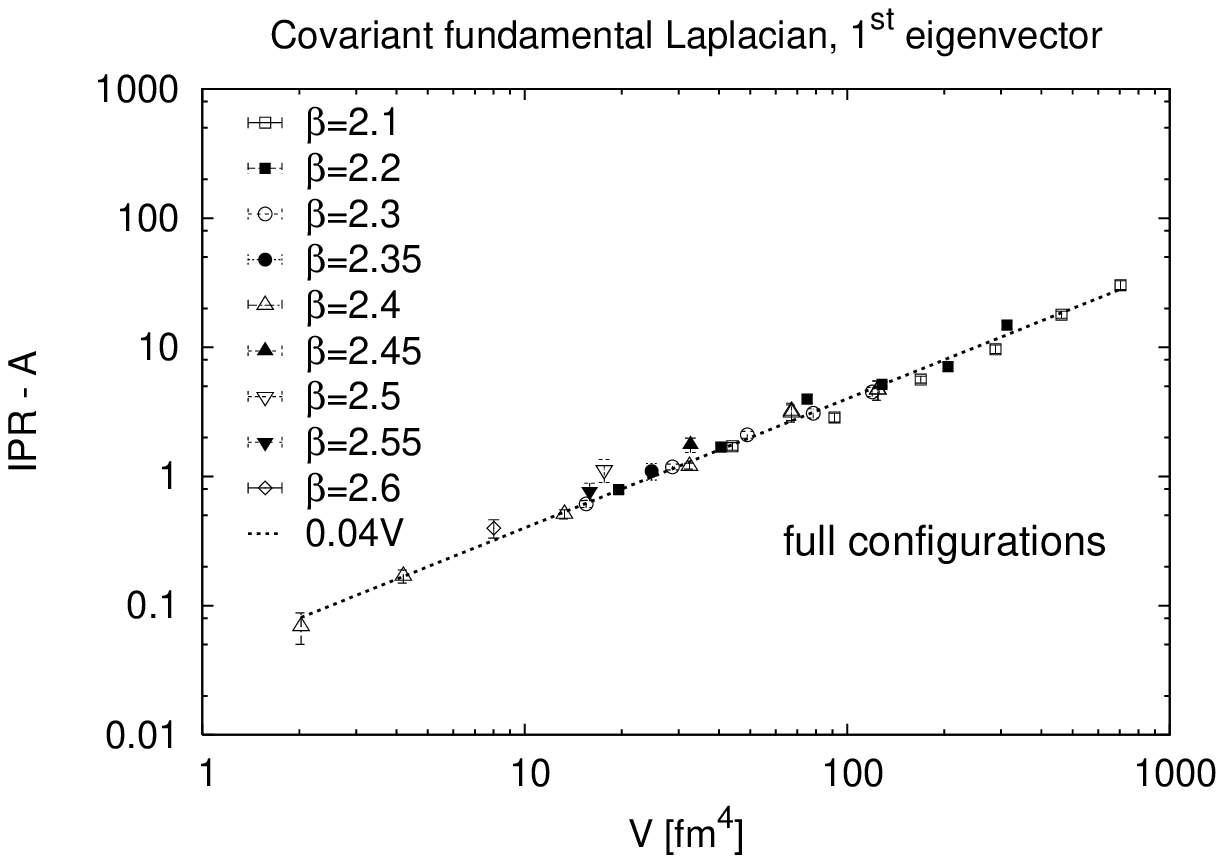}}}
\caption{Log-log plot of $IPR-A$ vs.\ lattice volume in physical units
(fm${}^4$). The dotted line is a best fit to linear dependence on
volume. }
\label{Vphys}
}

We have calculated the average IPR of the lowest-lying
eigenmode at $\b=2.1,2.2,2.3,2.4$, and
fit the data at each coupling to
\beq
        IPR(\b,L) = A(\b) + {L^4 \over b(\b)}
\label{fit}
\eeq
where $L$ is the lattice length in lattice units.
By the "subtracted" IPR data we refer to our data for $IPR(\b,L)$
minus the constant $A(\b)$ obtained from the best fit to eq.\
\rf{fit}. Figure \ref{IPR-fund} is a log-log plot of the
subtracted IPR vs.\  $L$, together with the best fit. The
proportionality of $IPR-A(\b)$ to lattice volume is quite clear
from the match of data points to the fit.
In Fig.\ \ref{Vphys} we plot the subtracted IPR data vs.\ the
lattice volume $V$ in physical units at a variety of couplings.
Here $V [\mbox{fm}^4] = V a^4(\b)=(La(\b))^4$ is the lattice
volume in physical units, where
$a=a(\b)$ denotes the lattice spacing (in fm) at coupling $\b$ .  In
these units we find that the IPR values at different $\b$ fall
roughly on the same straight line, which implies that the
localization 4-volume ($b a^4$) is constant in physical
units.\footnote{In this figure we
have included some IPR data at
$\b>2.4$ obtained on only a single volume: $L=24$ for
$\b=2.45,2.5$, and $L=28$ at $\b=2.55,2.6$.  In these cases, which
do not allow a fit to eq.\ \rf{fit}, we have subtracted
$A(\b)=1.38$ obtained at $\b=2.4$. Although these higher-$\b$ data
points are consistent with the multi-volume data obtained for
$\b\le 2.4$, they come with the caveat that the subtraction
constant $A(\b)$ was not obtained from a multi-volume fit.}

    The next question is whether the localization we have found
is due specifically to the confining disorder of the lattice
configuration, or whether it is due simply to local stochastic
fluctuations in the link variables.  To address this question, we
have computed the low-lying eigenmode spectrum in the
center-projected, and vortex-removed configurations.  Recall that
the vortex-removed configurations \cite{dFE} are obtained by
multiplying the lattice configuration in maximal center gauge by
the center-projected configuration.  This is known to remove the
confinement properties of the lattice configuration (the
asymptotic string tension is zero), but is otherwise a rather
minimal disturbance of the original lattice, since plaquette
values away from P-vortex locations are left unchanged.  Locally,
the vortex-removed configuration is actually somewhat more
disordered than the original configuration, since the plaquette
values at P-vortex locations are multiplied by $-1$.  Figure
\ref{vortex-only} shows that the lowest eigenmode computed in the
center-projected (or "vortex-only") configuration is localized
much as in the original lattice.  The vortex-removed data, however,
is drastically different.  In Fig.\
\ref{vortex-removed}, we see that that the IPRs of the vortex-removed
configurations remain essentially constant
as $L$ increases, and therefore localization is completely
absent.  The obvious
conclusion is that localization is associated with large-scale,
confining disorder provided by center vortices, rather than local,
non-confining fluctuations of the link variables.

\FIGURE[thb]{
\centerline{{\includegraphics[width=8truecm]{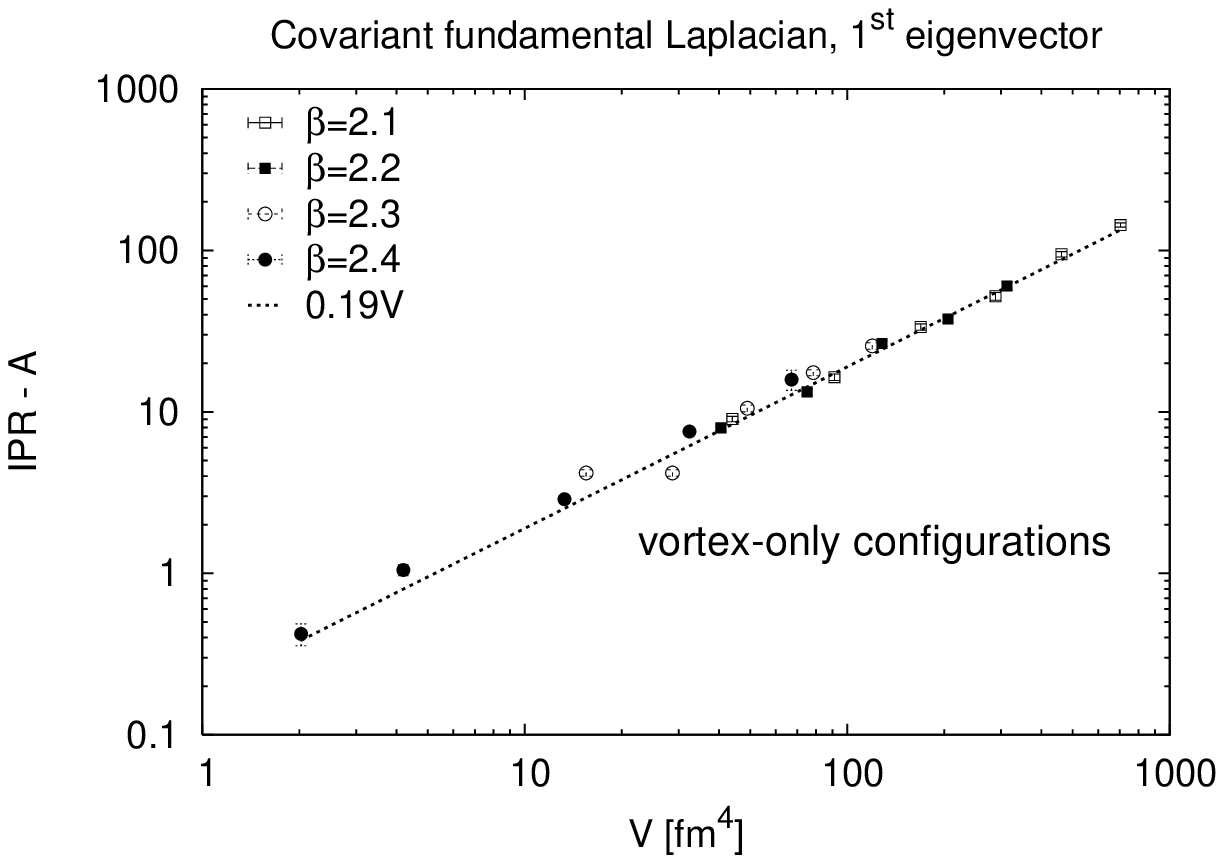}}}
\caption{Subtracted IPR of the lowest Laplacian eigenmode vs.\
lattice volume $V$ in physical units, for
the "vortex-only" (i.e.\ center-projected) lattice configurations.}
\label{vortex-only}
}

\FIGURE[thb]{
\centerline{{\includegraphics[width=8truecm]{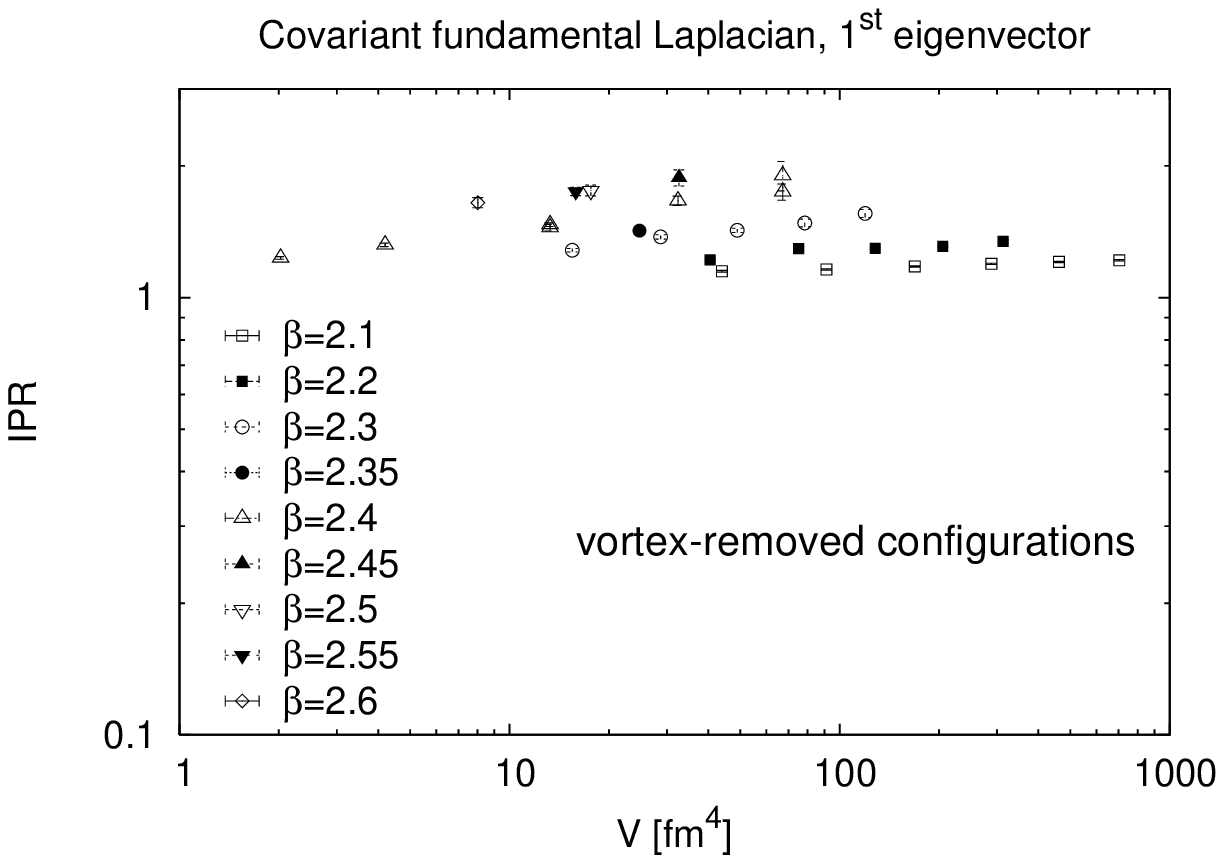}}}
\caption{Same as Fig.\ \ref{vortex-only} for the configurations with
vortices removed.  In this case the data is consistent with
extended, rather than localized, eigenmodes.}
\label{vortex-removed}
}

   The same conclusion is reached by examining the Remaining Norm
observable, defined in eq.\ \rf{RN}.  Figures
\ref{RN-fund}-\ref{RN-fund-nv} show the Remaining Norms at
$\b=2.1$ and at various lattice volumes, for unmodified,
center-projected, and vortex-removed configurations.  For the
unmodified configurations (Fig.\ \ref{RN-fund}), the curve does
seem to become slightly broader as the lattice volume increases,
although the indications are that the RN data converges to a
limiting curve at the largest volumes.  The convergence is much
clearer for the center-projected configurations, where almost all
the RN data falls on essentially the same curve (Fig.\
\ref{RN-fund-cp}).  For the vortex-removed data, there is no
convergence at all; the curve simply broadens as the volume
increases (Fig.\  \ref{RN-fund-nv}).\footnote{If we plot the
vortex-removed data vs.\ volume fraction rather than volume, all
the curves at different volumes coincide.}

\FIGURE[thb]{
\centerline{{\includegraphics[width=8truecm]{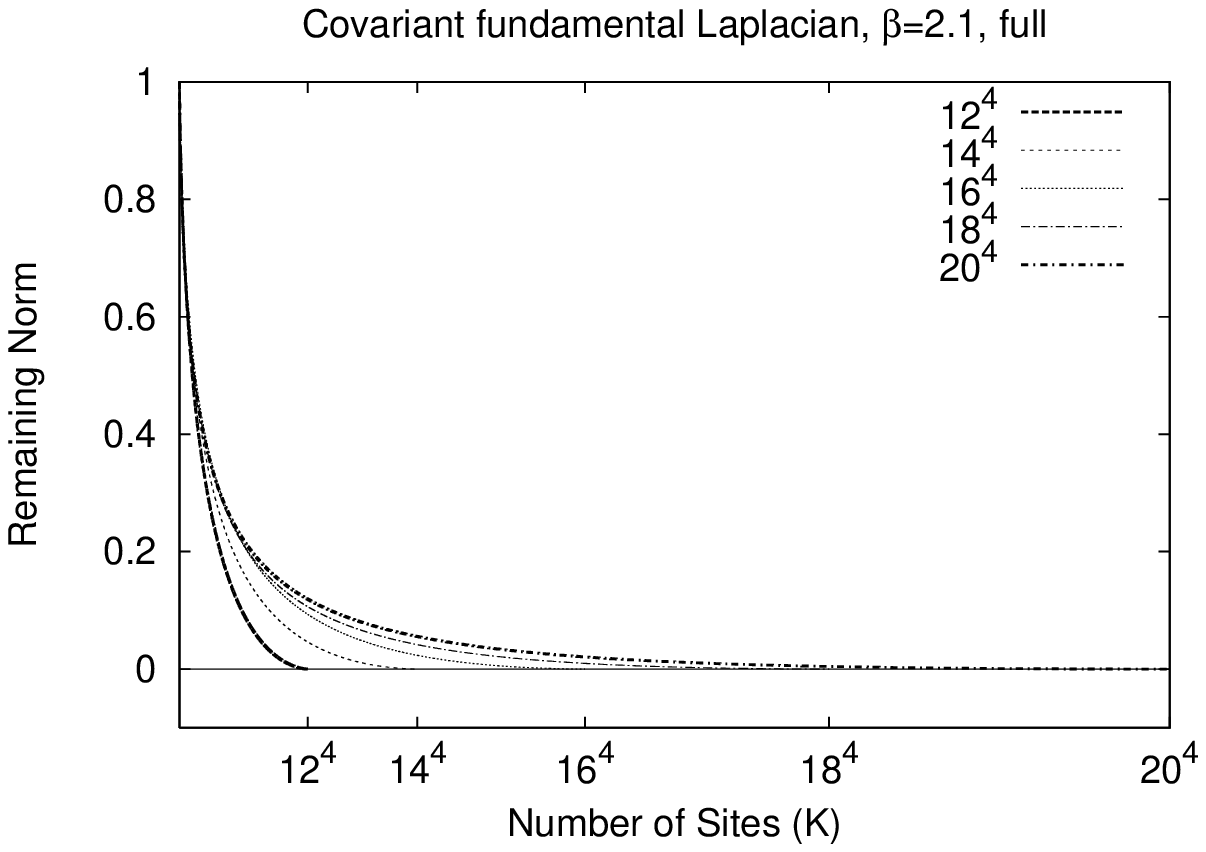}}}
\caption{Remaining Norm vs.\ volume at $\b=2.1$ for
unmodified lattices.}
\label{RN-fund}
}

\FIGURE[thb]{
\centerline{{\includegraphics[width=8truecm]{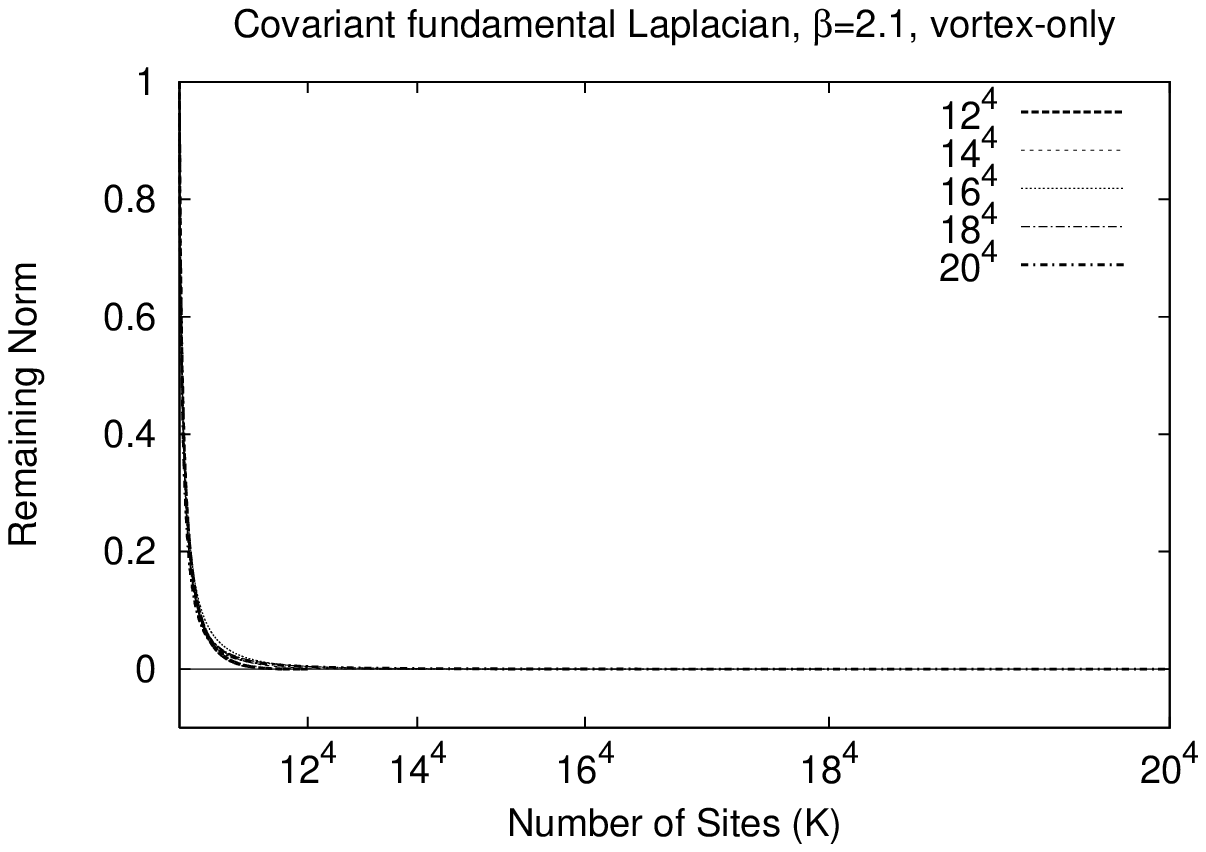}}}
\caption{Remaining Norm vs.\ volume at $\b=2.1$ for
center-projected lattices.}
\label{RN-fund-cp}
}

\FIGURE[thb]{
\centerline{{\includegraphics[width=8truecm]{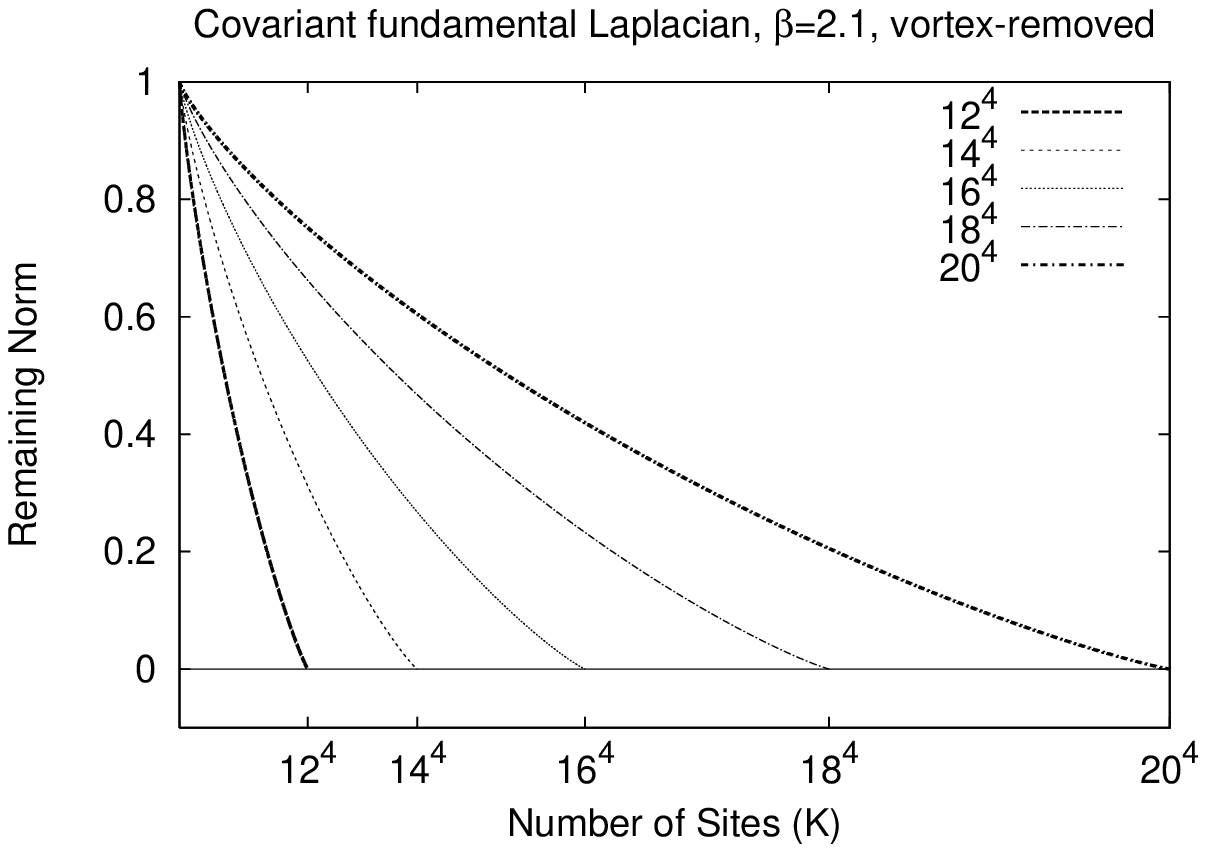}}}
\caption{Remaining Norm vs.\ volume at $\b=2.1$ for
vortex-removed lattices.}
\label{RN-fund-nv}
}

    Finally, in Fig.\ \ref{b-fund} we display the extension
$b^{1/4}a$ of the localized region in physical units, for both the
unmodified and the center-projected configurations, at couplings
$\b=2.1 - 2.4$.  The localization volume $b$, in lattice units, is
extracted from a best fit of the IPR data to eq.\ \rf{fit}. It
seems clear that the average extension of the localized region in
physical units is essentially $\b$-independent, with this length
roughly $2.3$ fm for unmodified configurations, and $1.5$ fm for
vortex-only configurations. In Fig.\ \ref{b-fund1} we display, for
individual lattice volumes at some higher $\b$ values, a quantity
which would equal (according to \rf{fit}) the inverse localization
volume $(ba^4)^{-1}$ in physical units. Again we observe from the
data that this quantity is insensitive to the value of $\b$ (with
the same caveats as in footnote 2 above).

\FIGURE[thb]{
\centerline{{\includegraphics[width=8truecm]{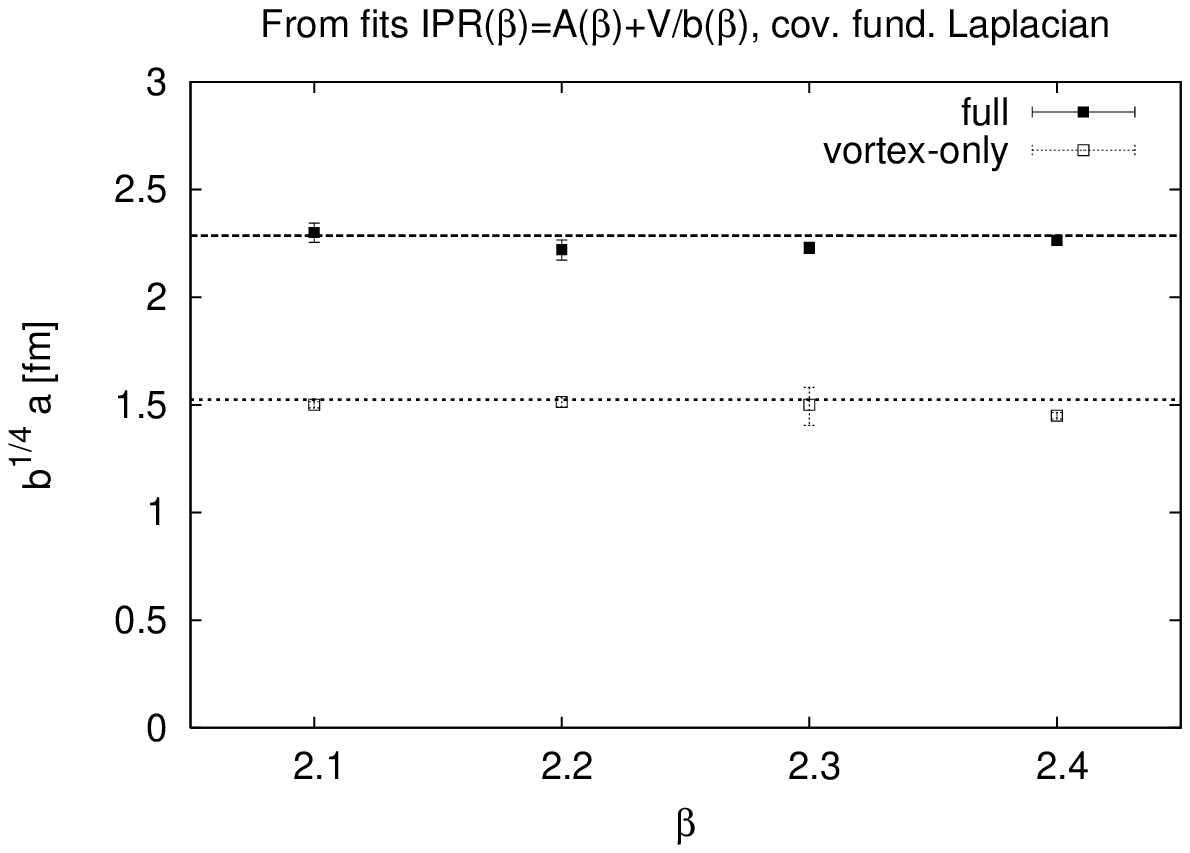}}}
\caption{Extension of the localization volume in physical units.}
\label{b-fund}
}

\FIGURE[thb]{
\centerline{{\includegraphics[width=8truecm]{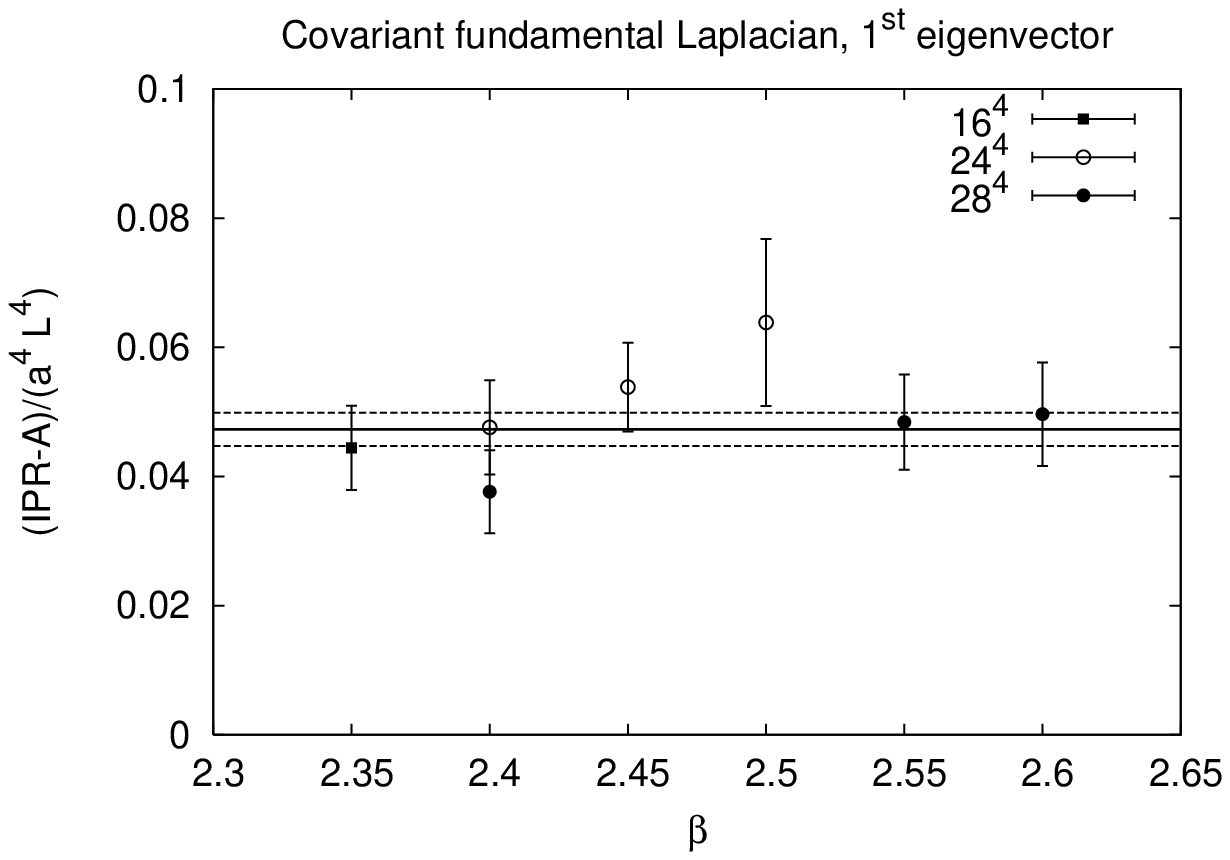}}}
\caption{Inverse localization volume in physical units (see text).}
\label{b-fund1}
}

   So far we have considered only the lowest eigenmode.
For the higher modes, we find at any fixed volume that the $IPR_n$
decreases ($b$ increases) as the eigenvalue $\l_n$ increases ,
until there is no localization at all.   For the non-localized
states, we see that the $IPR$ are small (O(1)) even on large
lattices, and do not increase with lattice size (Fig.\
\ref{mobedg}). A similar result is obtained in the
center-projected lattices (Fig.\ \ref{mobedg-nv}). This situation
is familiar in the Anderson model \cite{Anderson}, where
eigenmodes at the upper and lower ends of the spectrum are
localized, and eigenmodes in the bulk of the spectrum are
extended.  The boundaries between localized and extended states
are known as "mobility edges" \cite{Mott}.  In our case we observe
numerically a remarkable symmetry: If the eigenmodes $\p_n$ are
ordered from lowest ($n=1$) to highest $(n=n_{max}=2L^4)$
eigenvalue, then we find for \emph{each} lattice configuration
$U_\m(x)$ the correspondence
\beq
        IPR_n = IPR_{n_{max}-n+1}
\label{high-low}
\eeq
We emphasize that this identity holds for IPR's calculated in
each configuration, and not just for the expectation values.  The
identity follows from a symmetry.  Let $\phi^a_n(x)$ be an eigenmode
of the covariant Laplacian with eigenvalue
\beq
\l_n = 2D - \E_n
\eeq
where $-\E_n$ is the eigenvalue of the nearest-neighbor piece of
the Laplacian.  Then it is
not hard to see that
\beq
          \phi' = (-1)^{\sum_\m x_\m} \phi_n
\eeq
is also an eigenstate of the covariant Laplacian, with eigenvalue
\beq
          \l' = 2D + \E_n
\eeq Thus there is a pairwise correspondence of eigenstates with
eigenvalues below and above $\l=2D$, having exactly the same IPR
values. This accounts for the identity \rf{high-low}.

\FIGURE[thb]{
\centerline{{\includegraphics[width=8truecm]{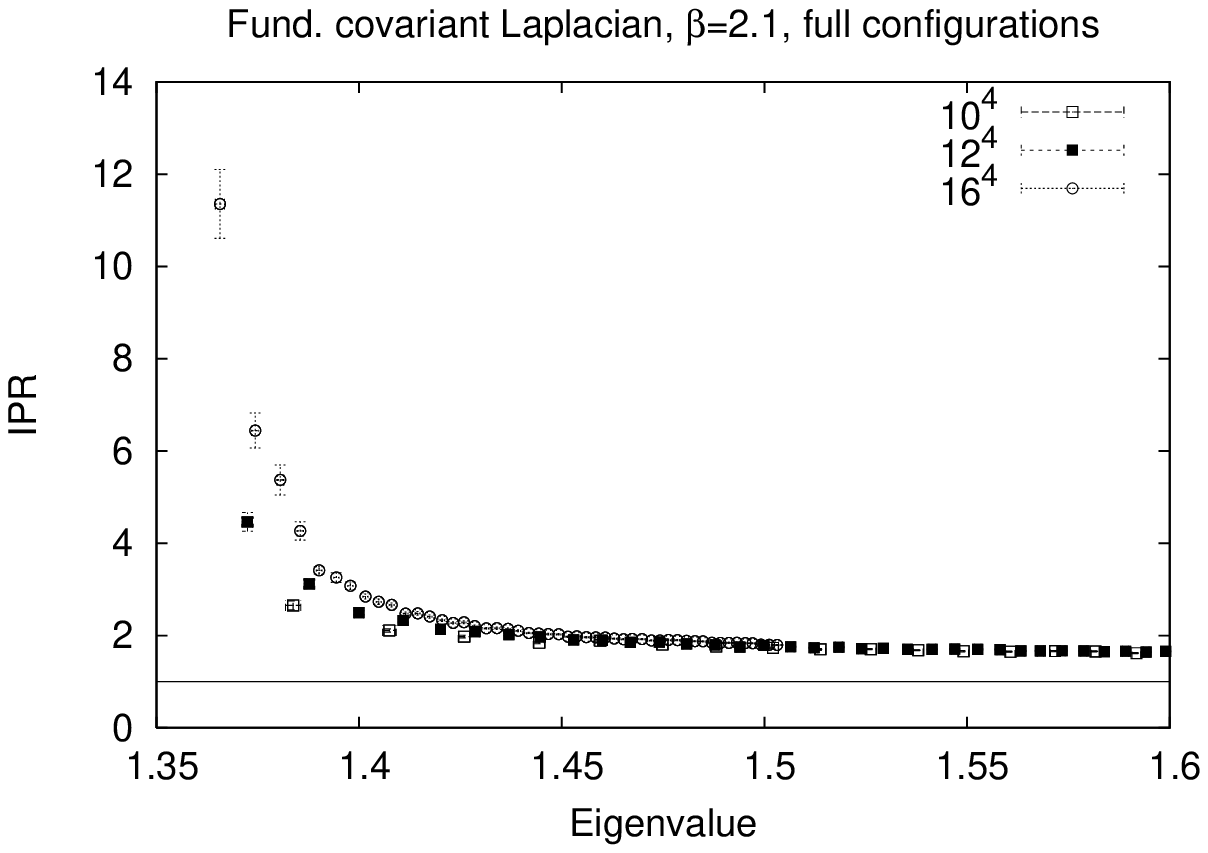}}}
\caption{IPRs of the first hundred eigenmodes of the covariant
Laplacian operator at $\b=2.1$.}
\label{mobedg}
}

\FIGURE[thb]{
\centerline{{\includegraphics[width=8truecm]{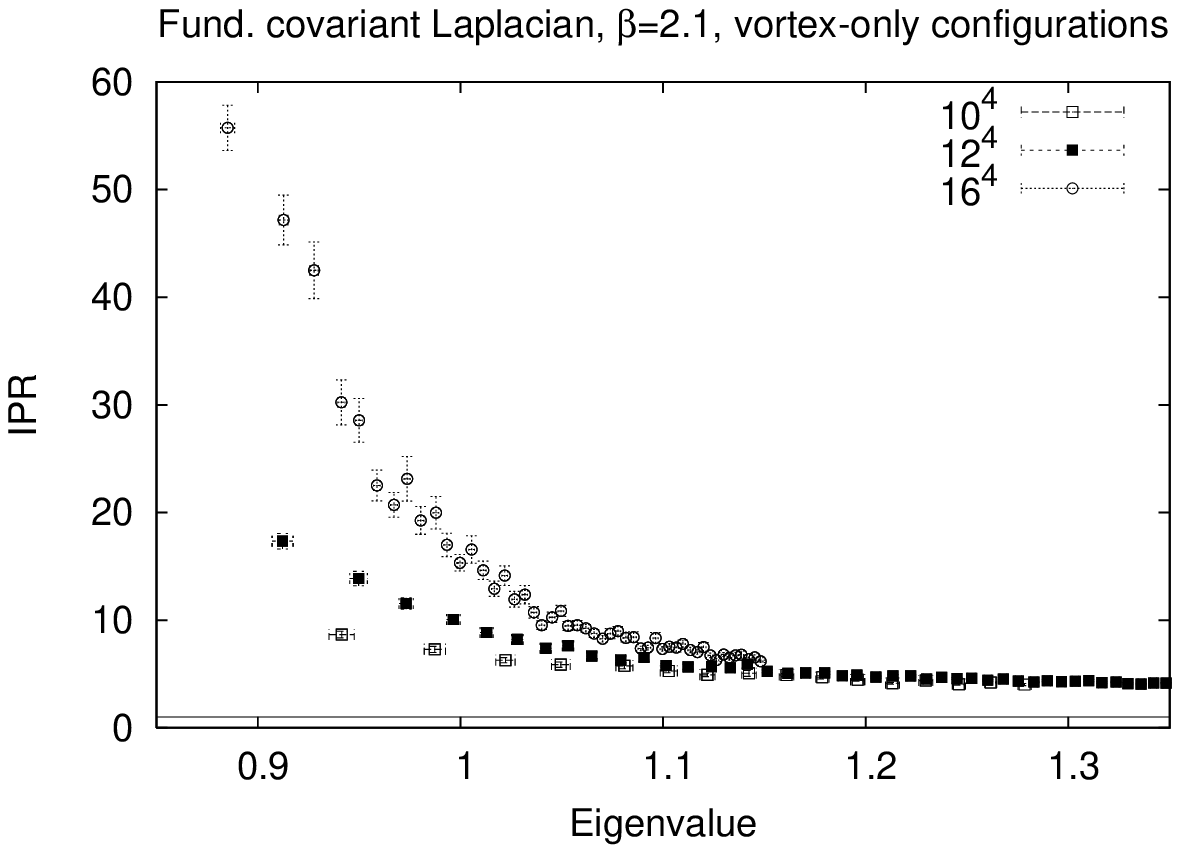}}}
\caption{Same as Fig.\ \ref{mobedg} for the center-projected
configurations.}
\label{mobedg-nv}
}

\section{Localization in the Adjoint Representation}

   One might expect that merely changing the gauge group representation
of the covariant Laplacian operator, from fundamental to adjoint,
would not greatly affect the localization properties of the
eigenmodes; certainly one does not expect to see any qualitative
difference.  Surprisingly, this expectation turns out to be
wrong.

   The covariant Laplacian operator in the adjoint representation of
SU(2) has the same form in as eq.\ \rf{triangle}, except that the
color indices run from 1 to 3, and the link matrices
\beq
         U^{ab}_\m = \oh \mbox{Tr}[\s^a U_\m(x) \s^b U^\dg_\m(x)]
\eeq
 are SO(3) group-valued (where the link variables inside the
trace are in fundamental representation). The results for the IPR
of the lowest eigenmode are shown in Figs.\ \ref{IPR-adj1} and
\ref{IPR-adj2}.  The first thing to notice is that in comparison
to the IPRs of the fundamental representation in Fig.\
\ref{IPR-fund}, the corresponding IPRs of the adjoint
representation are far higher. The difference, at fixed volume and
fixed coupling $\b$, is a factor of at least an order of
magnitude, and this factor grows with $\b$.   Moreover, in
contrast to the fundamental representation, the profile of the
Remaining Norm actually shrinks with increasing volume (cf.\
Figs.\ \ref{RN-adj} and \ref{RN-adj-all} for  $\b=2.3$),
indicating that the lowest eigenmode becomes, if anything,
\emph{more} localized as the lattice volume
increases.\footnote{The high degree of localization present in
low-lying adjoint representation eigenmodes was pointed out
several years ago by Bertle et al.\ \cite{Roman}, in connection
with a Laplacian version of direct maximal center gauge.}

\FIGURE[thb]{
\centerline{{\includegraphics[width=8truecm]{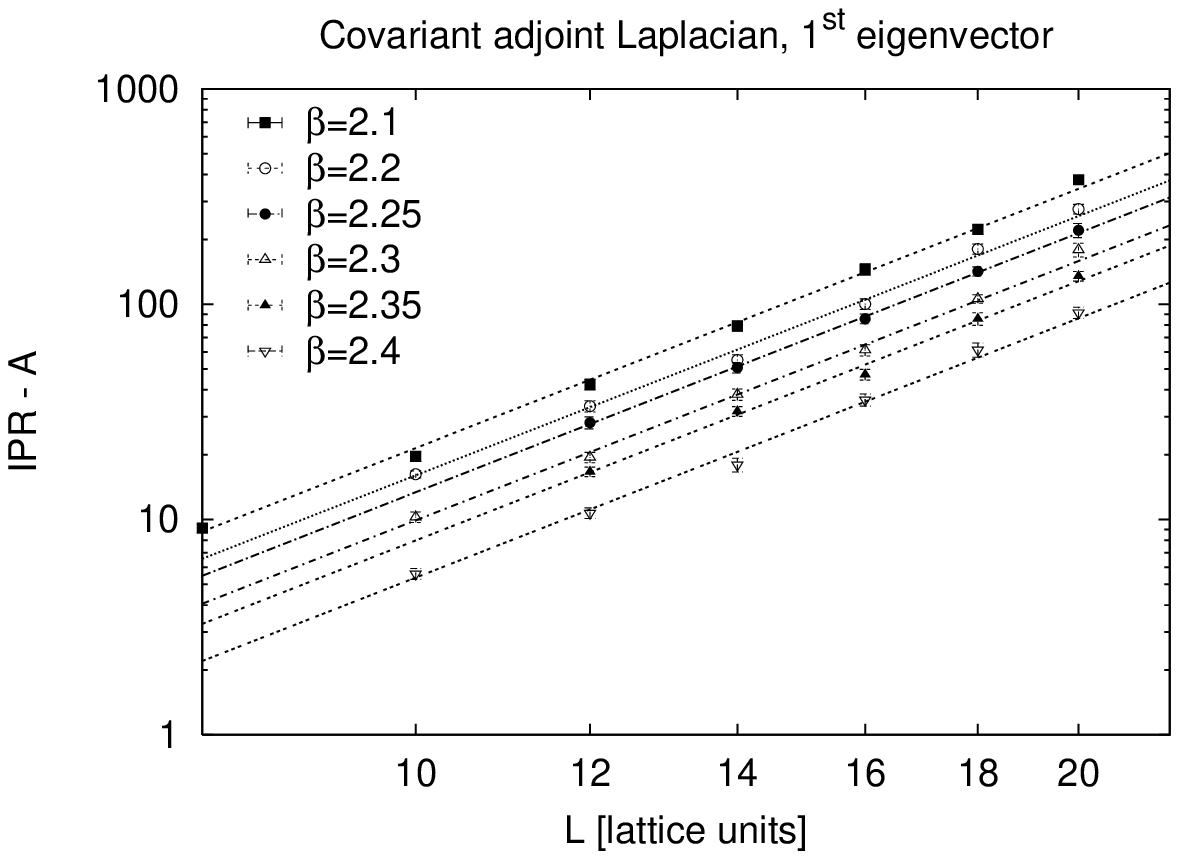}}}
\caption{Log-log plot of the subtracted IPR vs.\ lattice length $L$, for
the lowest eigenmode of the covariant Laplacian in the adjoint
representation, at $\b=2.1-2.4$.}
\label{IPR-adj1}
}

\FIGURE[thb]{
\centerline{{\includegraphics[width=8truecm]{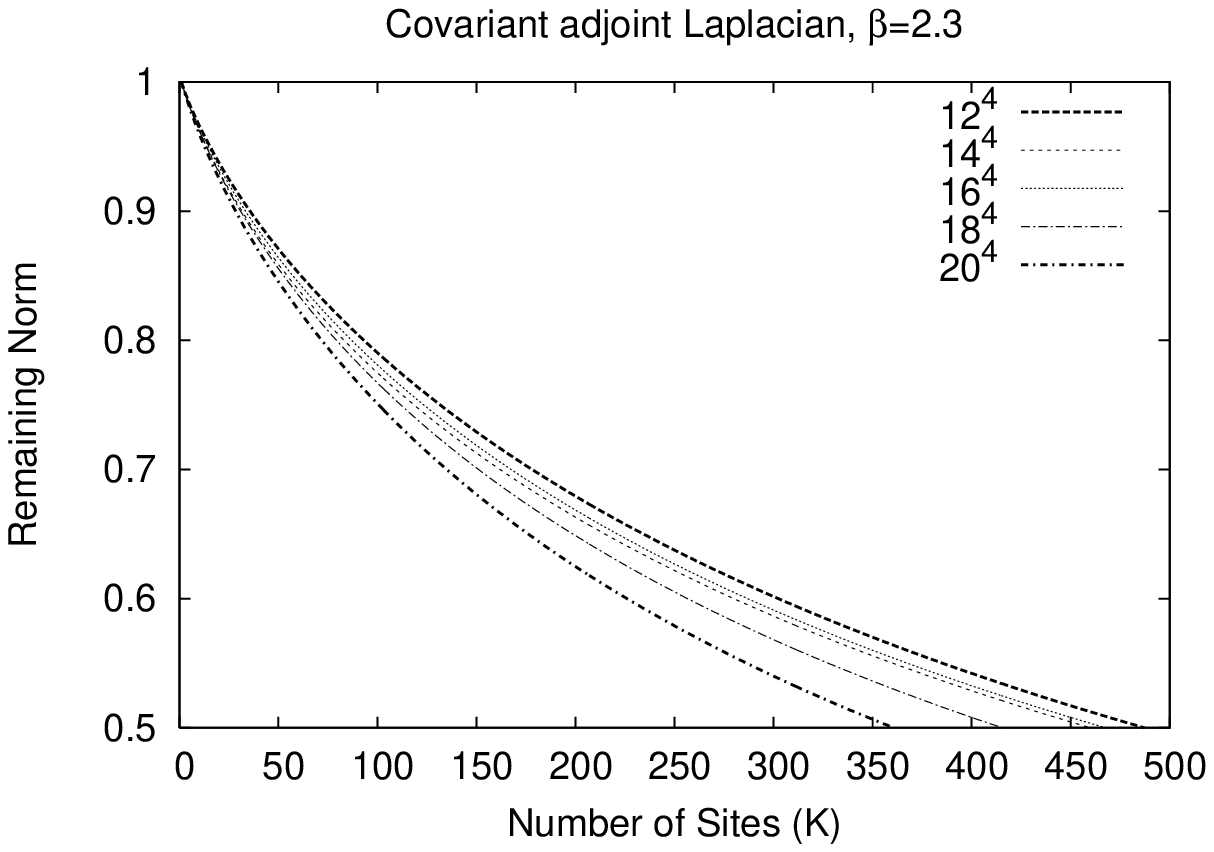}}}
\caption{Remaining Norm for the lowest eigenmode of the covariant
Laplacian, in the adjoint representation,
for various lattice volumes at $\b=2.3$.
Only the RN data for small numbers of sites are shown.}
\label{RN-adj}
}

\FIGURE[thb]{
\centerline{{\includegraphics[width=8truecm]{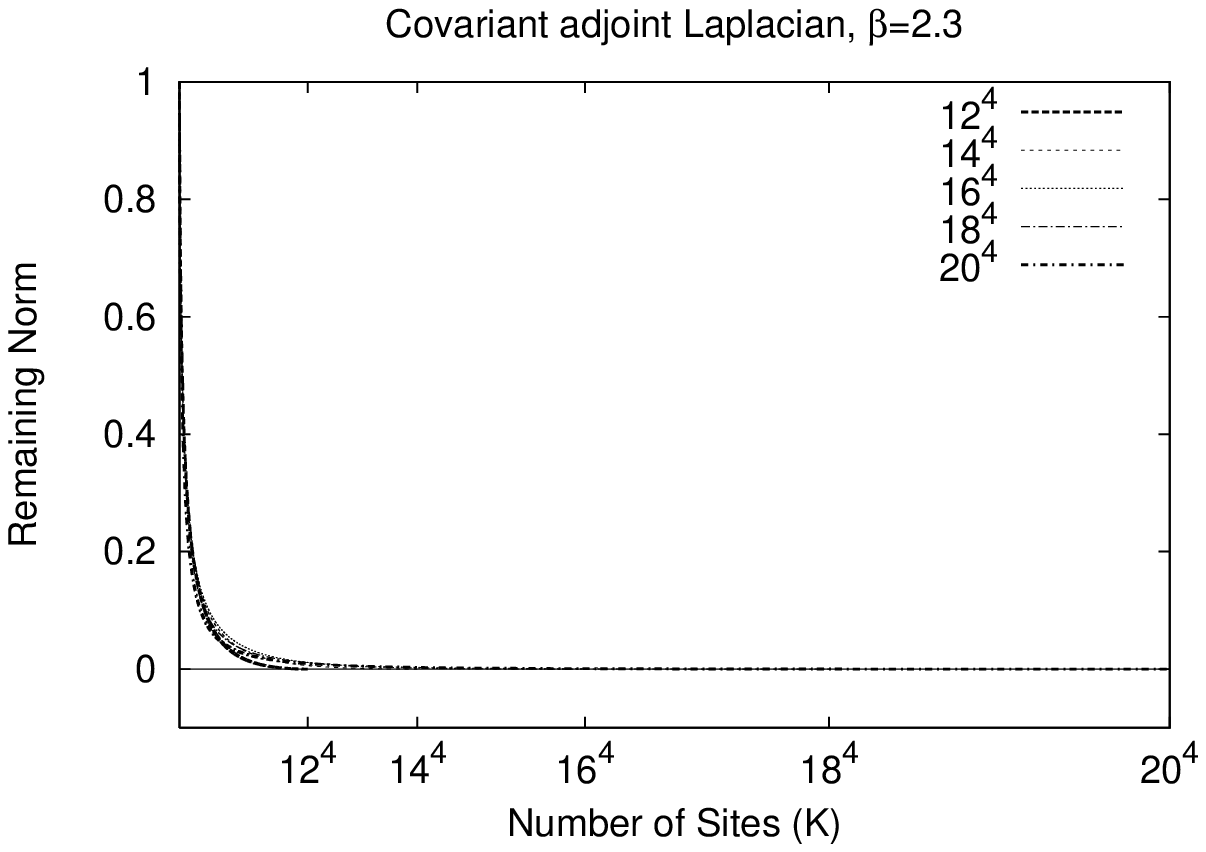}}}
\caption{Remaining Norm for the lowest eigenmode of the covariant
Laplacian, in the adjoint representation,
for various lattice
volumes at $\b=2.3$.  All RN data are shown.}
\label{RN-adj-all}
}

    However, if we extract $b(\b)$ from a fit to eq.\ \rf{fit},
we find that $ba^4$ falls rapidly with $\b$, indicating that the
localization volume shrinks in physical units.  Instead, if we
compute $ba^2$ at each $\b$, it is found that this quantity is
essentially constant.  In  Fig.\ \ref{IPR-adj2} we have plotted
$IPR \times a^2$ versus lattice volume in physical units, and it
is clear that the data points at different $\b$-values fall on the
same line.  This would imply, from eq.\ \rf{fit}, that $ba^2$
rather than $ba^4$ is $\b$-independent, which can be seen directly
in Fig.\ \ref{bsig}.  We also display, in Fig.\ \ref{IPR-adj3},
the data for $IPR/(a^2 L^4)$ at some  individual volumes at higher
$\b$. These points should approximate $1/(ba^2)$, and again the
data is only very weakly dependent on $\b$.

     The fact that $ba^2$ is constant, rather than $ba^4$, is
open to interpretation. It could be, following the suggestion in
ref.\ \cite{MILC}, that the lowest-lying eigenmode has its support
on a surface-like region of area $A_{phys}$, whose expectation
value is $\b$-independent, and lattice scale thickness. If the
thickness is, for example, only a single lattice spacing, then in
physical units
\beq
         b a^4 = A_{phys} a^2
\eeq
Thus, if $b a^2$ is $\b$-independent as $\b\ra \infty$, that
means that the surface area $A_{phys}$ has a finite extension in
physical units.  This in turn implies that the lowest eigenmode of
the covariant Laplacian, in adjoint representation, has its
support on a 2-brane of finite extension.  Assuming minimal (one
lattice spacing) thickness, one can estimate from
$\s_{phys}=(440~\mbox{MeV})^2$ and Fig.\ \ref{bsig} that the area
of this brane is $A_{phys} \approx 26~\mbox{fm}^2$.  This
possibility is attractive, because of the possible connection with
ultraviolet renormalons \cite{Valya} and/or P-vortex surfaces
\cite{ITEP}.

\FIGURE[thb]{
\centerline{{\includegraphics[width=8truecm]{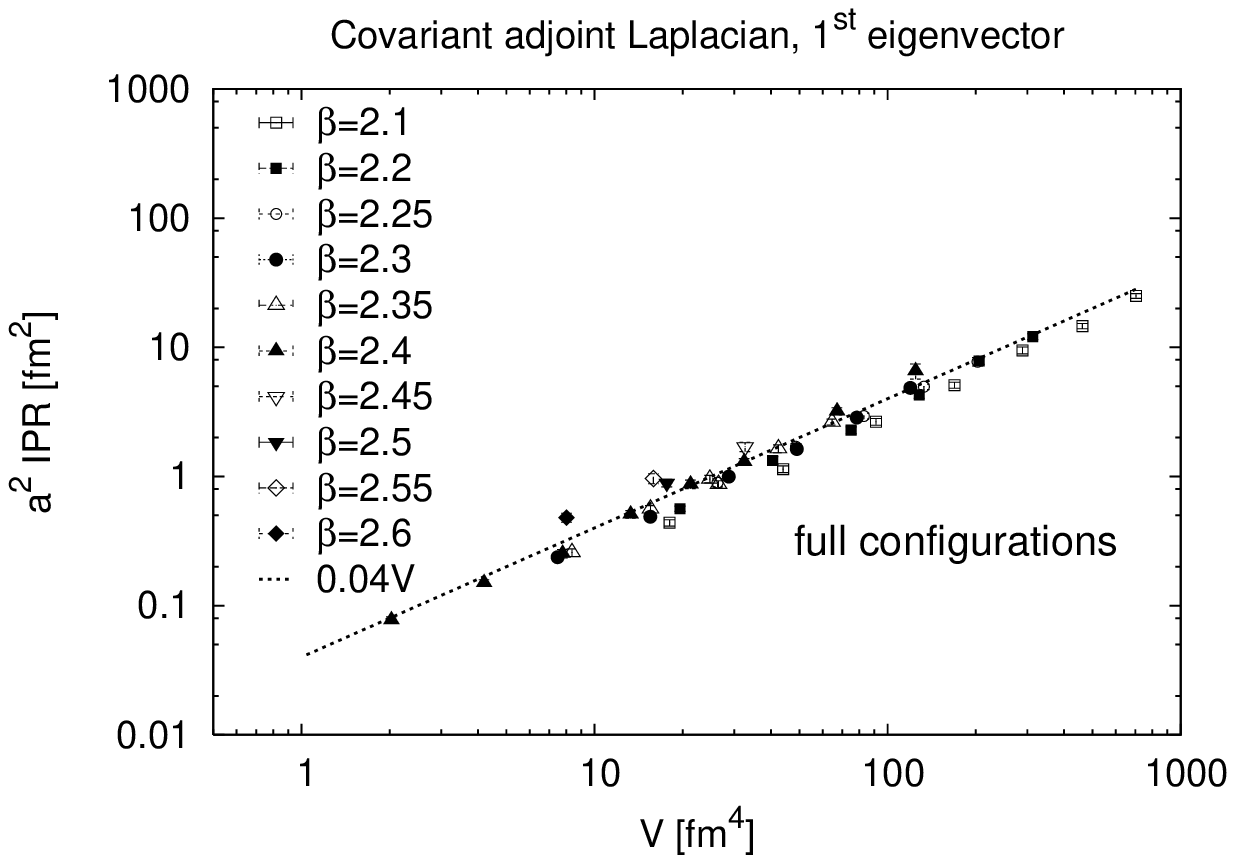}}}
\caption{Log-log plot of adjoint representation $IPR\times a^2$
vs.\ lattice volume in physical units (fm${}^4$). }
\label{IPR-adj2}
}

\FIGURE[thb]{
\centerline{{\includegraphics[width=8truecm]{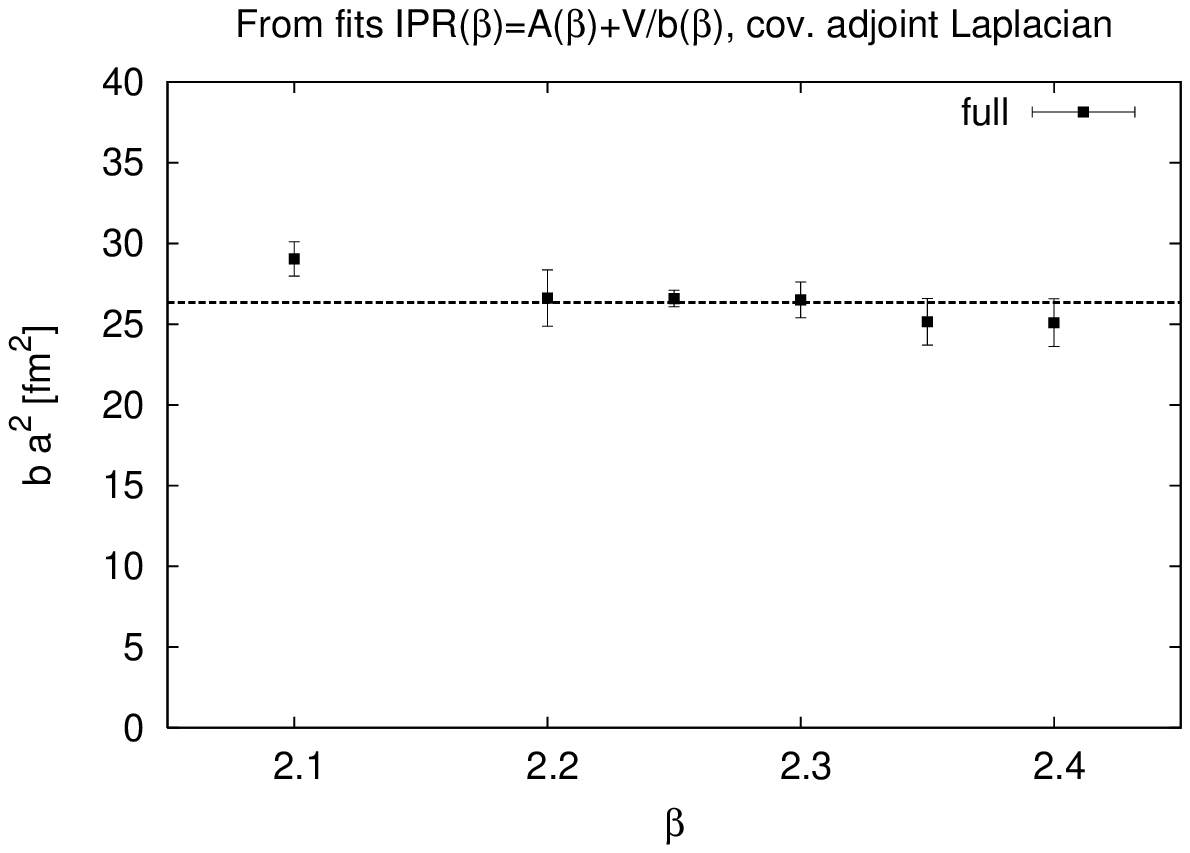}}}
\caption{$ba^2$ in the range
$\b \in [2.1,2.4]$.  If $b a^2$ is a constant at large $\b$,
then the localization volume shrinks to zero, in physical units,
in the continuum limit.}
\label{bsig}
}

\FIGURE[thb]{
\centerline{{\includegraphics[width=8truecm]{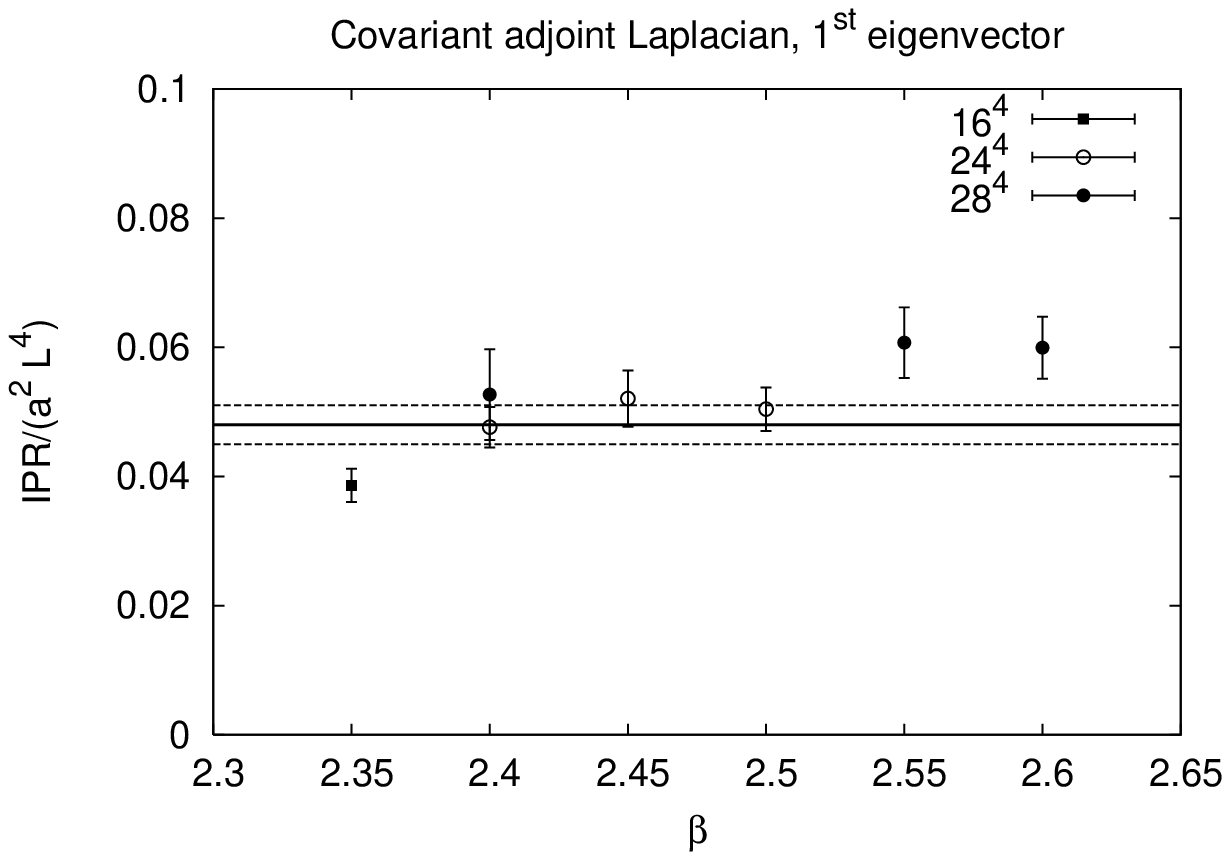}}}
\caption{Plot of $IPR/(a^2 L^4)$ at some individual volumes at
higher $\b$ values.}
\label{IPR-adj3}
}

   However, a more straightforward interpretation of Fig.\
\ref{bsig} is that the localization region is a 4-volume whose
extension simply shrinks to zero in physical units, in all
directions, in the continuum limit.  To test which of these
two possibilities is correct, we define the rms extension
$\D x$ of the localization region as follows:  On a given
lattice, let $x_0^\m$ be the site at which the density
$\r(x)$ is largest, and let
\beq
         d_\m(x) = \left\{ \begin{array}{cl}
                     |x^\m - x_0^\m| & \mbox{if}~ \le L/2 \cr
                L - |x^\m - x_0^\m|  &  \mbox{otherwise} \end{array}
                \right.
\eeq
Then
\beq
        \D x^2 = \left\langle \sum_x (d_1^2+d_2^2+d_3^2+d_4^2) \r(x)
                 \right\rangle
\eeq
If the localization region were surface-like, with finite surface area
in physical units, then unless the surface is space-filling
we expect that $\D x$ is finite in physical units.  On the other
hand, if the eigenmode has its support on a 4-volume of extension $\D x$ in all
directions (i.e.\ ``point-like''), then we expect that that the
volume of a 3-sphere of radius
$\D x$ (volume $= \oh \pi^2 \D x^4$) would be comparable to the
localization volume $b\approx V/IPR$.  In fact, computing the ratio
\beq
       \eta = { V/IPR \over \oh \pi^2 (\D x)^4}
\eeq
we find that $\eta \approx 0.5\pm 0.15$, over the whole range of
$\b \in [2.1,2.4]$.  We conclude that the localization region is
in fact point-like rather than surface-like, with a volume that
shrinks to zero in physical units in the continuum limit.

   The point-like nature of the lowest eigenmodes is confirmed by viewing
the eigenmode density $\r(x)$
for typical thermalized lattices.
We have made computer animations of $\r(x)$,
on a series of two and three volume slices of the four-dimensional lattice,
available on the web at the URLs cited in ref.\ \cite{animate2} (2D slices)
and \cite{animate3} (3D slices).

\subsection{The Higgs Phase}

    The next question is whether the sharp localization of the lowest-lying
eigenmode in adjoint representation is due to
large-scale confining disorder, or whether it is merely due to
local gaussian fluctuations at the lattice scale.
Since the adjoint representation is insensitive to the
center of the gauge group, it is not possible to cancel out the
confinement disorder in this representation by the de
Forcrand-D'Elia trick.  Instead we can add a Higgs field, and
investigate what happens to localization as we adjust the Higgs
coupling to go from the confinement-like to the Higgs phase.  For
the SU(2) gauge group, and the Higgs field in the fundamental
representation, the action can be written as \cite{Lang}
\beq
    S = \b \sum_{plaq} \oh \mbox{Tr}[UUU^\dg U^\dg]
      + \gamma \sum_{x,\m} \oh \mbox{Tr}[\phi^\dg(x) U_\m(x)
\phi(x+\widehat{\m})]
\eeq
where $\phi(x)$ is an SU(2) group-valued field.
It is known that (i) this theory has no
local, gauge-invariant order parameter which can signal a phase
transition; (ii) there is no confinement asymptotically for any $\g>0$;
and (iii) any two
points in the $\b - \gamma$ phase diagram can be joined by a line
along which the free energy is entirely analytic \cite{FS}.
There is nonetheless a real distinction between
the Higgs and the "confinement-like" phases.  If we fix to Coulomb
gauge, then there remains on any time-slice a global remnant of
the gauge symmetry.  This remnant symmetry is unbroken in the
"confinement-like" phase, and broken in the Higgs phase
\cite{GOZ}.\footnote{The symmetry-breaking order parameter is non-local
when expressed as a gauge-invariant quantity, so there is no contradiction
of the Fradkin-Shenker-Osterwalder-Seiler \cite{FS} result.} Physically, this implies
that an isolated charge,
together with its Coulomb field, is a finite energy state in the
Higgs phase, and infinite energy in the confinement-like phase.

   At $\b=2.1$, the transition from the confinement-like to the
Higgs phase occurs at around $\gamma=0.9$.  In Fig.\
\ref{IPR-higgs} we display the IPRs at $\g=0.7$ (confinement-like
phase) and $\g=1.2$ (Higgs phase).  The drastic reduction in IPR
in the Higgs phase as compared to the confinement phase is
evident.  This does not mean that in the Higgs phase the magnitude
of the lowest eigenmode is homogeneous throughout the lattice
volume, but we do find that the lowest eigenmode becomes
substantially more extended as the lattice volume increases.
The qualitative difference in lowest eigenmodes of the two phases
is clear from a plot of the remaining norms at $\b=2.1$, shown for the
confinement-like phase ($\g=0.7$) in  Fig.\ \ref{RN-hg07}, and
the Higgs phase ($\g=1.2$) in Fig.\ \ref{RN-hg12}.

\FIGURE[thb]{
\centerline{{\includegraphics[width=8truecm]{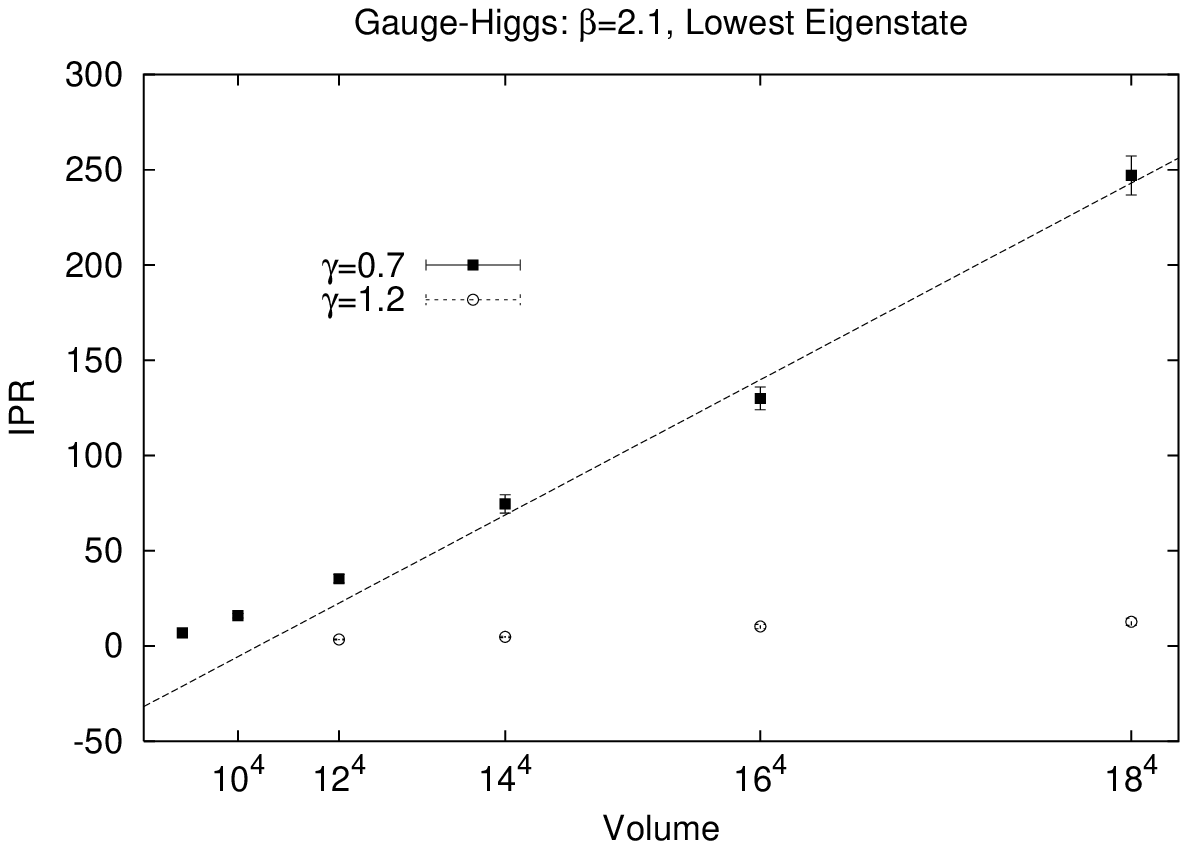}}}
\caption{IPRs in SU(2) gauge-Higgs theory at $\b=2.1$
and $\g=0.7$ (confinement-like phase) and $\g=1.2$
(Higgs phase).}
\label{IPR-higgs}
}

\FIGURE[thb]{
\centerline{{\includegraphics[width=8truecm]{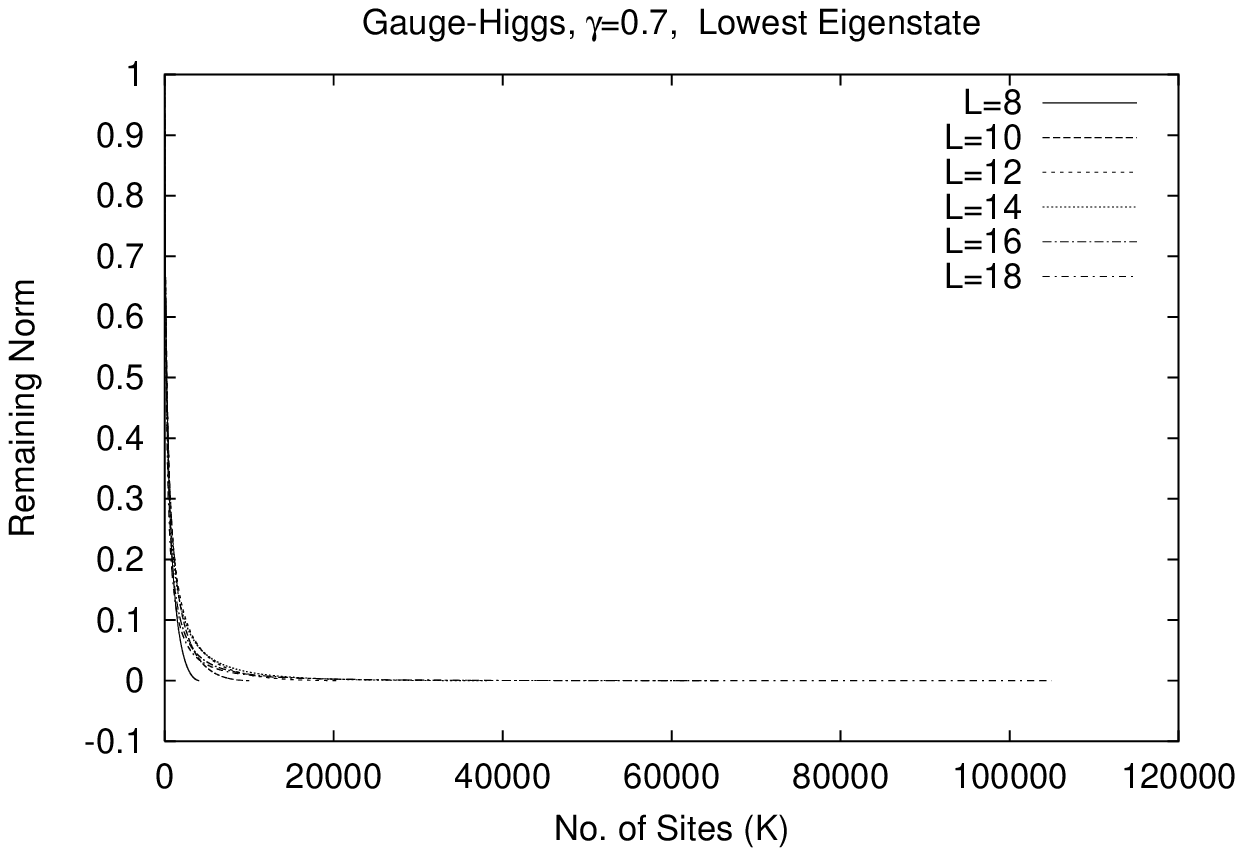}}}
\caption{Remaining norms in the confinement-like phase
on $L^4$ lattices, $L=8-18$.}
\label{RN-hg07}
}

\FIGURE[thb]{
\centerline{{\includegraphics[width=8truecm]{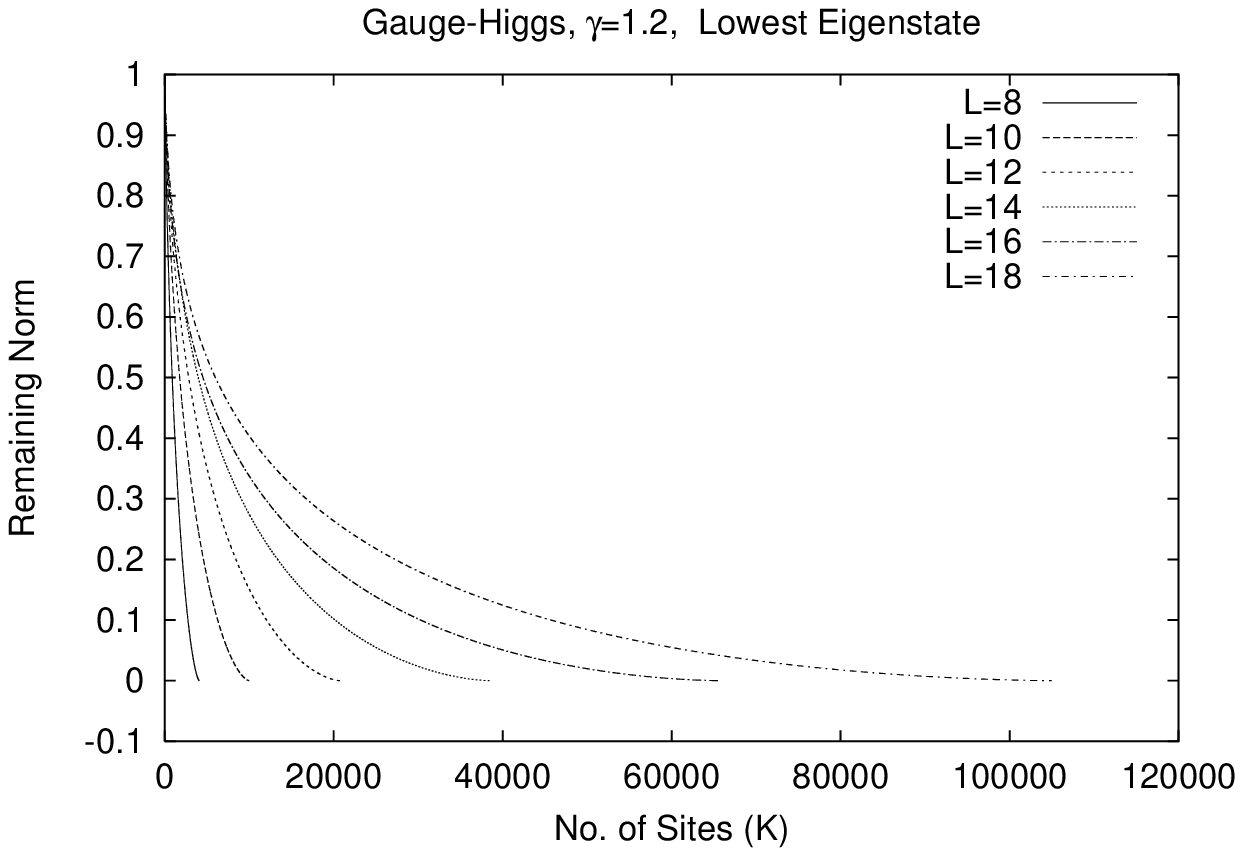}}}
\caption{Same as Fig.\ \ref{RN-hg07}, for the Higgs phase.}
\label{RN-hg12}
}

   In order to express these differences a little more quantitatively,
we note that in an Anderson localized state $\rho(x)$ decays exponentially
away from the localization region.  This means that the sum
\beq
        s_p = \sum_x \r^{1/p}(x)
\eeq
is finite in the infinite volume limit, for any $p>0$.  In contrast, if
$\rho(x)$ has only a power-law falloff (or no falloff at all, for homogeneous
states), then there will be some $p>1$ such that $s_p \ra \infty$ in the
infinite volume limit.  In Fig.\ \ref{s2} we display our results for $s_2$,
computed at $\g=0.7$ and $\g=1.2$.  In the confinement-like phase, $s_2$
is constant at approximately $s_2 = 71$, while in the Higgs phase, $s_2$
appears to rise linearly, on $L^4$ lattices, with extension $L$.  This
confirms that the lowest eigenmode of the adjoint Laplacian in the Higgs
phase is not an exponentially localized Anderson-type state.

\FIGURE[thb]{
\centerline{{\includegraphics[width=8truecm]{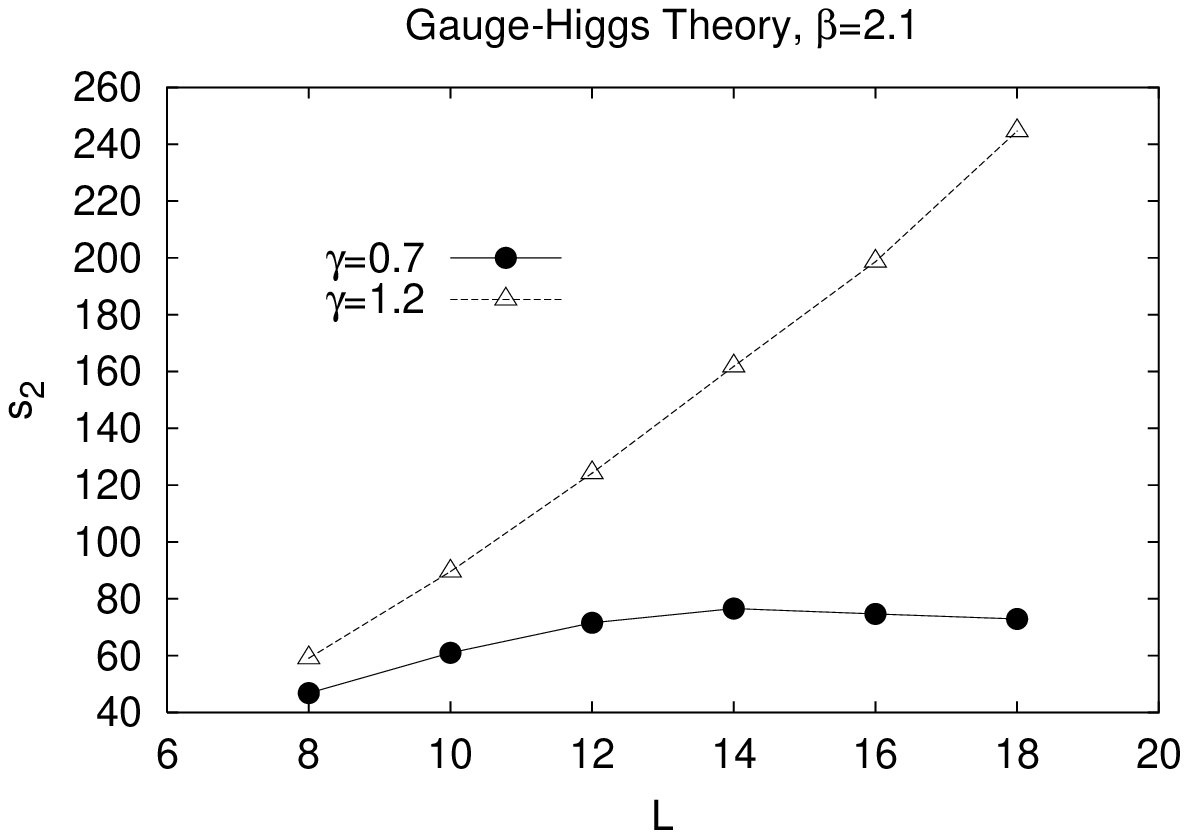}}}
\caption{Test of Anderson (exponential) localization: $s_2$
vs.\ volume in the Higgs and confinement-like phase.}
\label{s2}
}

\section{The $\bf j=3/2$ Representation}

   We have also calculated the low-lying eigenmode spectrum of the
covariant Laplacian in the $j=3/2$ representation of the SU(2)
gauge group, at various volumes and $\b$ values.
We find that the low-lying modes are
even more strongly localized in the $j=3/2$ representation than in the adjoint
representation, and the data falls on a single line when plotting
$IPR\times a^4$ vs.\ lattice volume $(La)^4$ in physical units.
This data suggests that $b$ itself is approximately $\b$-independent,
and if this feature remains true at higher $\b$, it implies that the
localization region is always on the scale of the lattice spacing,
regardless of the length of the lattice spacing in physical units.

The $j=3/2$ representation of SU(2) has the same n-ality
as the fundamental representation.  However, unlike
the fundamental representation, vortex removal has essentially no
effect on localization of the low-lying $j=3/2$ eigenmodes, as seen by
comparing the no-vortex result in Fig.\ \ref{b} with the previous
Fig.\ \ref{a}.  Given the high degree of localization in the $j=3/2$
representation, insensitivity to vortex removal is not surprising.
It is rather obvious that eigenmodes of center-projected configurations,
in the $j=3/2$ representation are in complete correspondence to the
eigenmodes of the $j=1/2$ representation, and would therefore be
localized on the scale of fermis.  This sort of physical-scale localization,
which is removed in the fundamental representation by vortex removal, cannot
easily account for the lattice-scale localization found in the $j=3/2$
representation.

\FIGURE[thb]{
\centerline{{\includegraphics[width=8truecm]{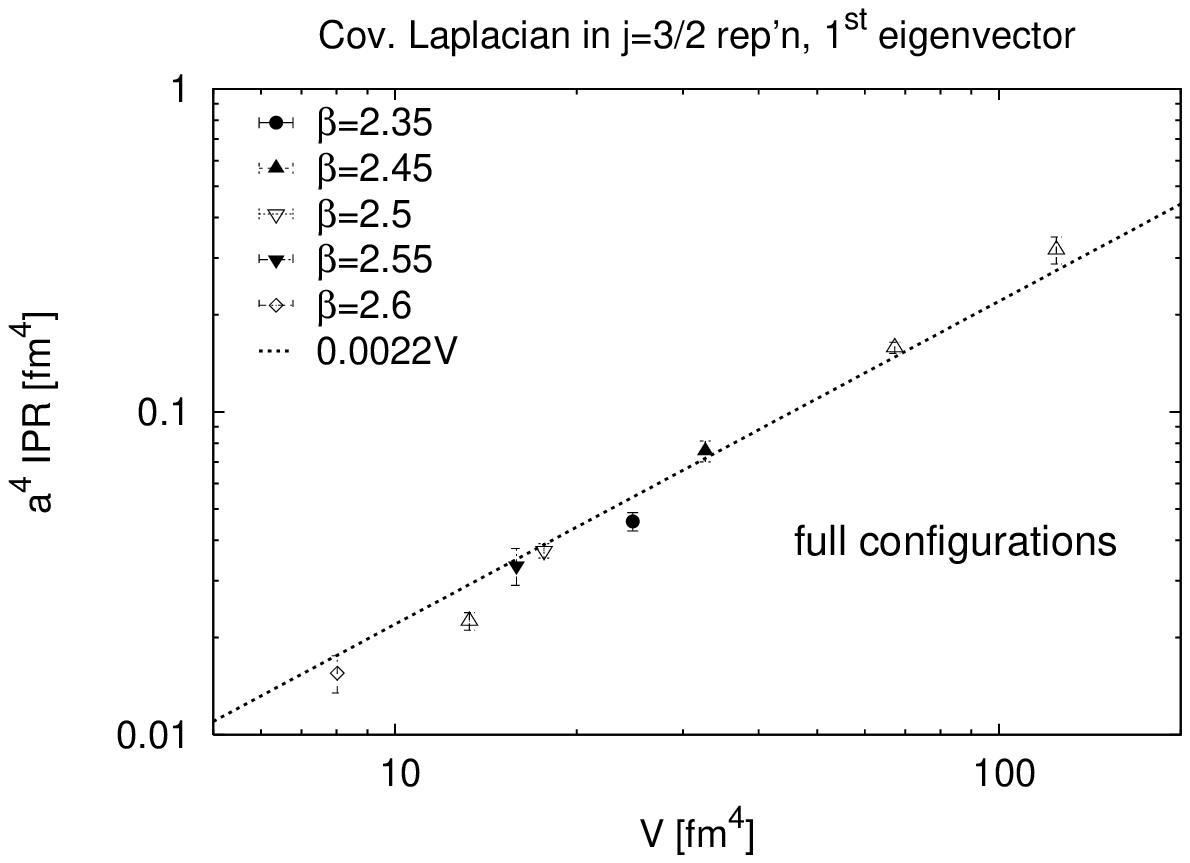}}}
\caption{Log-log plot of $IPR \times a^4$ vs.\ lattice volume in physical
units, for the $j=3/2$ representation.}
\label{a}
}

\FIGURE[thb]{
\centerline{{\includegraphics[width=8truecm]{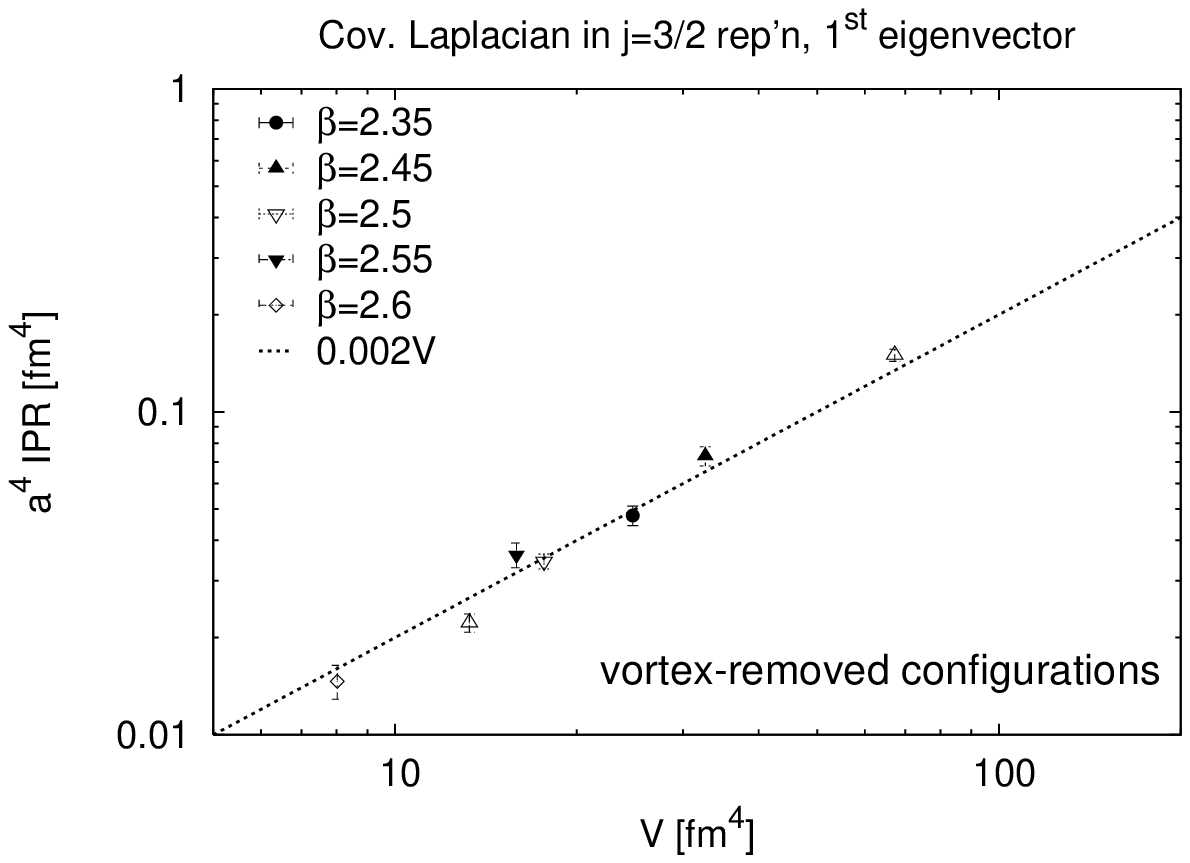}}}
\caption{Same as Fig.\ \ref{a}, for vortex-removed configurations.}
\label{b}
}

\section{Localized Modes vs.\ Massless Particles}

    The propagator of a color-charged scalar particle of
bare mass $m_0^2$, propagating in a background lattice gauge
configuration $U_\m(x)$, can be expressed as
\bea
      S^{ab}(x,y;U) &=& \left[ {1\over -\triangle + m_0^2}
                           \right]^{ab}_{xy}
\non \\ \non \\ \non \\
                    &=& \sum_n {\phi_n^a(x) \phi_n^{\dg b}(y) \over \l_n + m_0^2}
\eea
where the $\p_n^a(x)$ are the eigenmodes of
of the covariant lattice Laplacian operator.
The operator $S^{ab}$ is gauge covariant, and therefore its
expectation value vanishes in the absence of gauge-fixing,
regardless of confinement properties.  If a physical (e.g.\
Coulomb) gauge is chosen, then we expect that $\langle
S^{ab}(x,y;U)\rangle = 0$ in a confining theory if $x_0\ne y_0$,
because a color-charged state should not be able to propagate.
This requirement is enforced in Coulomb gauge, as explained in
ref.\ \cite{GOZ}, by the existence of an unbroken remnant global
gauge symmetry, which ensures that $S^{ab}$ will still average to
zero for finite time separations.

   However, suppose we wish to restrict our discussion to
gauge-invariant operators.  Then the next simplest two-point
function to consider is
\beq
       K(x-y)  =  \langle S^{ab}(x,y,U) S^{ba}(y,x,U) \rangle
\eeq
which can be regarded as the quenched approximation to the VEV
of two identical scalar fields $\vph^{1,2}$
\beq
  \Bigl\langle \vph^{1a}(x) \vph^{\dg 2a}(x) \vph^{1b}(y)
      \vph^{\dg 2b}(y) \Bigr\rangle
\eeq where the two "flavors" are introduced just to avoid two
scalars annihilating into an intermediate state of gauge bosons.

   Let us now ask whether the bare mass $m_0^2$ can be adjusted
so as to make $K(x-y)$ long range.  If that is possible, then the
two scalars must couple to an intermediate state containing only
massless particles. These massless intermediate particles might be
\begin{enumerate}
\item  The scalars themselves.  In a non-confining theory such as
QED${}_4$, it is generally possible to adjust $m_0^2$ so that the
physical mass $m_{phys}$ of the scalar particles takes on any
value desired; in particular $m_{phys}=0$.
\item  A two-scalar
bound state $B$, with $m_B=0$. \item  "Gluelump" states $G$.
Scalars in the adjoint representation can bind to gluons to form
colorless bound states.  In that case, it might be possible to
adjust $m_0^2$ so that $m_{G}=0$.
\end{enumerate}

   We will argue that localization of the low-lying states rules
out any long-range, power-law falloff (such as $1/|x-y|^4$ or
$1/|x-y|^2$) in $K(x-y)$, at least if tachyonic modes are
excluded, and hence the mass of the scalar particles cannot be
adjusted to zero as in non-confining theories. The reasoning is
closely related to that of McKane and Stone \cite{McKane} (see
also Golterman and Shamir \cite{Maarten}), in connection with
Anderson localization and the absence of massless Goldstone
particles.

The argument goes as follows: We can express $K(x-y)$ as a double
mode sum
\beq
       K(x-y) = \left\langle \sum_n {\p^a_n(x) \p^{\dg b}_n(y)
                \over \l_n + m_0^2}
          \sum_m {\p^b_m(y) \p^{\dg a}_m(x) \over \l_m + m_0^2} \right\rangle
\label{product}
\eeq
Suppose that we could adjust $m_0^2$ such
that $m_{phys}$ or $m_G$ were massless.  Then the two-point
function $K(x-y)$ is long-range, only falling off like
$1/|x-y|^4$, so that
\bea
         {1\over V} \sum_{x} \sum_{y} K(x-y) &=&
 \left\langle {1\over V}\sum_n {1\over (\l_n + m_0^2)^2} \right\rangle
\non \\
    &\ra& \int_{\l_{min}}^{\l_{max}} d\l ~ {\r(\l) \over (\l+m_0^2)^2 }
\label{mode_sum}
\eea
is logarithmically divergent at volume $V\ra
\infty$ (where $\r(\l)$ is the normalized density of eigenvalues).
It is assumed that there are no tachyonic modes; i.e.\ $m_0^2 \ge
-\l_{min}$.  Then from \rf{mode_sum} we see that (i) an infrared
divergence is possible only if $m_0^2 = -\l_{min}$; and (ii) if
there is a divergence, then it is due entirely to the low-lying
eigenmodes of the covariant Laplacian.  It is therefore sufficient
to restrict our attention to the near-zero modes of $-\triangle +
m_0^2$. However, if the near-zero modes are localized, then it is
impossible for those modes to be responsible for a long range in
$K(x-y)$.  Remove those modes, and the sum over the remaining
modes in eq.\ \rf{mode_sum} is finite, so a long range can't be
coming from the higher modes either.  The conclusion is that
$K(x-y)$ is short-range.  Given localization of the low-lying
eigenmodes, it is just not possible to adjust $m_0^2$ such that
intermediate states of mass $m_B,~m_G,$ or $m_{phys}$ become
massless, at least not without introducing tachyonic modes into the
kinetic operator $-\triangle + m_0^2$.

   There is still one puzzle.  The rhs of eq.\ \rf{mode_sum} can
certainly be infinite if $m_0^2 = -\l_{min}$, but how is it
possible for the lhs to diverge if $K(x-y)$ is short range? The
answer is that if we have Anderson localized states, the infinity
in \rf{mode_sum} is not coming from the \emph{sum} over $x,y$,
because in fact there is a divergence at \emph{any} $x,y$.  Notice
that the operator inside brackets in eq.\ \rf{product} contains a
piece (from $n=m$)
\beq
       G(x,y) = \sum_n {\p^{\dg a}_n(x) \p^a_n(x)
                        \p^b_n(y) \p^{\dg b}_n(y)
                \over (\l_n + m_0^2)^2}
\eeq
and note that the argument of the sum is positive.  Now suppose first
that all states $\p_n$ are extended, and normalize to unity.  Then
$|\phi_n(x)| \sim 1/\sqrt{V}$, which means that the numerator
contributes a factor of $1/V^2$.  This has to be compensated by
sums over $x,y$, contributing a factor of $V^2$, and then the
divergence is determined by the integral
\beq
       {1\over V} \sum_n {1\over (\l_n + m_0^2)^2}
          \ra \int_{\l_{min}}^{\l_{max}} d\l {\r(\l) \over (\l + m_0^2)^2}
\eeq
If we take $V\ra \infty$ without summing over positions, then
$G(x,y) \ra 0$.  So in this case, the divergence (if it exists) is
infrared in origin.  By contrast, suppose the near-zero modes of
$(-\triangle + m_0^2)$ are localized.  Presumably the number of
these modes grows linearly with volume.  For each $x,y$, there is
one particular state $\phi_n$ which gives the largest contribution
to the mode sum, and from this contribution
\beq
       G(x,y) \sim {1\over b^2} {\exp[-{(|x-z_n}|+|y-z_n|)/L]
           \over (\l_n + m_0)^2}
\eeq
where $|\phi_n(x)|$ is maximized at $x=z_n$, and $b$ is the
localization volume.  Certainly $G(x,y)$ may diverge if $\l_n +
m_0^2 \ra 0$, but this does not imply any divergence in the range
of $G(x,y)$ in the same limit.

\subsection{Eigenmodes of the Faddeev-Popov Operator}

   An interesting check of the preceding logic is provided by
the Coulomb gauge Faddeev-Popov (F-P) operator.\footnote{We thank
Dan Zwanziger for this suggestion.} In the continuum, this
operator looks superficially much like the covariant Laplacian in
D=3 dimensions, i.e.
\beq
       M = - \vec{\nabla} \cdot \vec{D}(A)
\eeq where $D_k(A)$ is the covariant derivative, and $A$ is fixed
to Coulomb gauge.  On the lattice, with SU(2) link variables
expressed as
\beq
      U_\m(x) = b_\m(x) + i \vec{a}_\m \cdot \vec{\s}
\eeq
the F-P operator becomes
\bea
 M^{ab}_{xy} &=& \d^{ab} \sum_{k=1}^3\Bigl\{ \d_{xy}  \left[b_k(x)
   + b_k(x-\hat{k})\right]
\non \\
   &-& \left.  \d_{x,y-\hat{k}} b_k(x)
   -  \d_{y,x-\hat{k}} b_k(y) \right\}
\non \\
      &-& \epsilon^{abc} \sum_{k=1}^3\left\{ \d_{x,y-\hat{k}} a^c_k(x)
                 - \d_{y,x-\hat{k}} a^c_k(y)  \right\}
\label{M}
\eea

   Because of its strong similarity to the covariant Laplacian, one
might expect that the low-lying eigenmodes of the F-P operator
would also be localized.  However, there is a strong argument
against this.  In Coulomb gauge, the Coulomb energy due to a
static color charge distribution $\r^a(x)$ is given by
\beq
       H_{coul} = \oh \int d^3x d^3y\;\J^{-\oh}\r^a(x) \J
               K^{ab}(x,y;A) \r^b(y) \J^{-\oh}
\eeq
where
\bea
       K^{ab}(x,y;A) &=& \left[ M^{-1}
       (-\nabla^2) M^{-1} \right]^{ab}_{xy}
\non \\
       \r^a &=& \r_q^a - g f^{abc} A^b_k E^{{\rm tr},c}_k
\non \\
        \J &=& \det[-\nabla \cdot D(A)]
\eea
In the scenario for confinement advocated by Gribov and
Zwanziger \cite{Gribov}, $\langle K^{ab}(x,y,A) \rangle$ rises
linearly with separation at large $|x-y|$, and this has been
confirmed by numerical simulation \cite{coulomb}.  But this
behavior can never come about if $M^{-1}$ is short range.  It
follows, from the previous arguments, that if $M^{-1}$ is
long-range then the low-lying eigenmodes of the F-P operator must
be extended, rather than localized. The F-P operator actually has
three trivial (constant) zero modes with $IPR=1$. Fig.\
\ref{IPR-FP} shows the IPR vs.\ volume of the lowest non-trivial
eigenmode at $\b=2.1$.  The IPR values, even on sizable lattices,
are rather low, and the growth in IPR with volume is very
gradual. The low IPRs imply that the lowest non-trivial mode is an
extended state (at least on the lattice sizes shown), despite the
similarity of the F-P operator to the covariant Laplacian, but in
accordance with the reasoning presented above.

\FIGURE[thb]{
\centerline{{\includegraphics[width=8truecm]{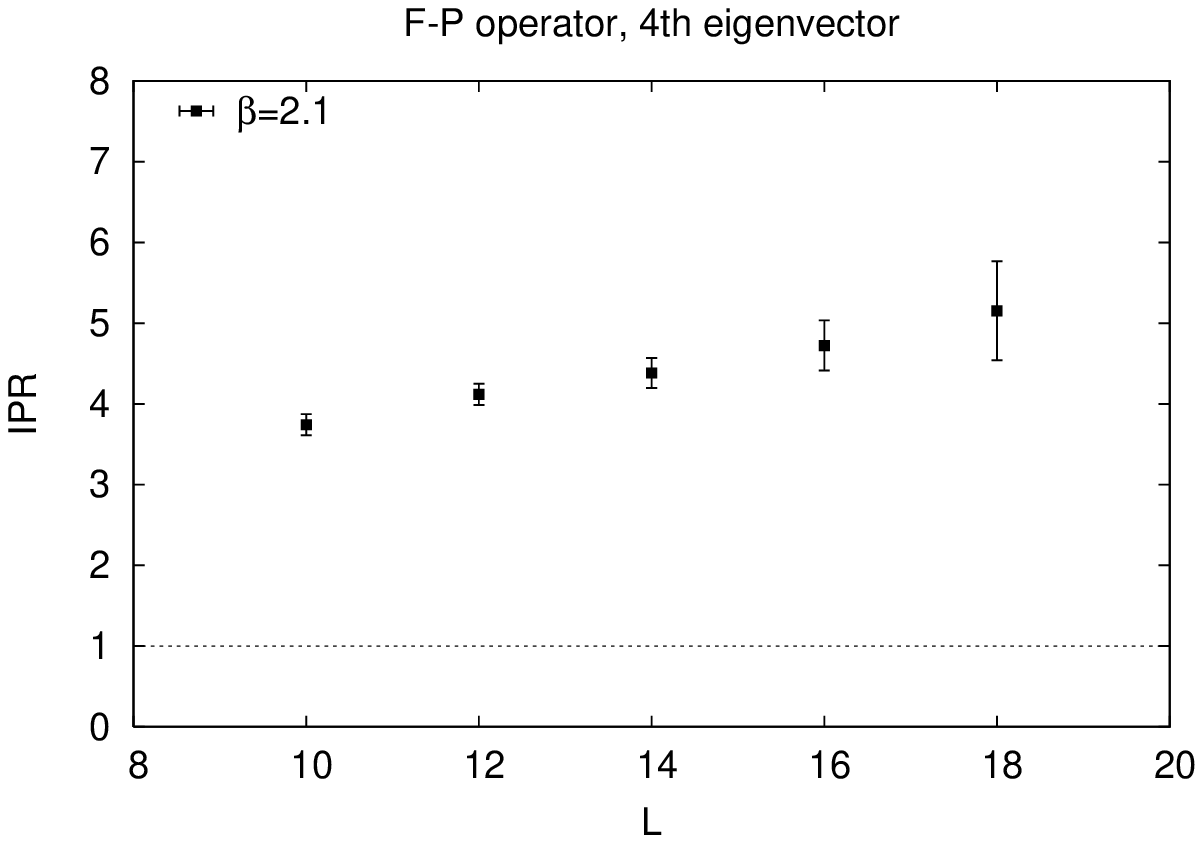}}}
\caption{IPR of the lowest non-trivial eigenmode of the
Faddeev-Popov operator.}
\label{IPR-FP}
}

\section{Conclusions}

   The low-lying eigenmodes of kinetic operators on the lattice,
such as the Dirac operator and the covariant Laplacian, are probes
which may be useful in exploring the properties of the QCD vacuum.
In this article we have begun a study of the low-lying eigenmodes
of the Yang-Mills covariant Laplacian, and have uncovered some
intriguing properties of those eigenmodes which we cannot as yet
explain.

    The most puzzling feature concerns gauge-group representation
dependence.  We have found that eigenmodes of the Yang-Mills
covariant Laplacian operator are localized at the lower and upper
ends of the spectrum, and that the degree of localization of these
eigenmodes depends \emph{qualitatively} on the group
representation of the Laplacian operator.  With $b$ denoting the
localization volume, in lattice units, of the lowest-lying
eigenmode, the data indicates that:
\begin{enumerate}
\item In the fundamental representation, the localization volume
$b(\b) a^4(\b)$ in physical units is
$\b$-independent, with an average extension on the order of 2 fm.
The localization property disappears
when center vortices are removed from the lattice configuration,
while the localization property of center-projected lattice
configurations is similar to that of unmodified
configurations.
\item  In the adjoint representation, the quantity $b(\b) a^2(\b)$
is $\b$-independent.  This means that at weak couplings the
localization volume is large in lattice units, but goes to zero,
in the continuum limit, in physical units.  The adjoint
representation is insensitive to vortex removal, but we find that
exponential localization, in a gauge-Higgs theory, goes away in
the Higgs phase.
\item  In the $j=3/2$ representation, $b$ itself is insensitive
to coupling, which indicates that the localization volume is on
the scale of the lattice spacing.  Vortex removal has no effect on
this high degree of localization.
\end{enumerate}

   We have also found that localization is absent in the eigenmodes
of the Faddeev-Popov operator in Coulomb gauge, despite the close
resemblance of the F-P operator to the covariant Laplacian.  This
absence of localized, low-lying F-P eigenmodes is implied by the
existence of a confining color-Coulomb potential, and is therefore
essential in Coulomb gauge confinement scenarios.

    Our results cast some doubt on the
utility of the so-called Laplacian gauges \cite{Uwe}, particularly
on large lattices.  Laplacian versions of various gauges
\cite{lap_gauges} have been introduced in the literature; these
have the advantage of avoiding the Gribov copy problem.  However,
each of these gauges makes use of either the lowest, or the lowest
two or three Laplacian eigenmodes, on the grounds that these are
uniquely the smoothest of the eigenmodes.  We have found,
however, that the lowest eigenmodes are actually the \emph{least}
smooth of all the Laplacian eigenmodes, and on a very large
lattice are negligibly small on most of the lattice volume.
Presumably, such highly localized modes are not well motivated for
the purpose of gauge-fixing.

The relationship of localization to confinement deserves further
investigation. In the fundamental representation, where
localization volume is finite in physical units, there is evidence
of a strong connection between confinement, center vortices and
the localization effect.  The relationship between localization
and confinement disorder is not so clear for the higher group
representations, where the localization lengths tend to zero in
the continuum limit, and are unaffected by vortex removal.  It is
possible that localized Laplacian eigenmodes in different group
representations are probing different features, and perhaps
different length scales, of the Yang-Mills vacuum state.

%
\acknowledgments{%
We have benefited from discussions with P. van Baal, M. Golterman, I. Horv\'ath, A. Larkin,
P. Marko{\v{s}}, Y. Shamir, O. Sushkov,
and D. Zwanziger. Our research is supported in part by the U.S.
Department of Energy
under Grant No.\ DE-FG03-92ER40711 (J.G.), the Slovak Science
and Technology Assistance Agency under Contract No.\  APVT-51-005704 (\v{S}.O.), grants RFBR 04-02-16079,  RFBR 05-02-16306-a (M.P.),  grants RFBR 05-02-16306-a, RFBR 05-02-17642, and
an Euler-Stipendien (S.S.).

%
%

\end{document}